\documentclass[iop]{emulateapj}
\submitted{Received March 27; accepted ???}

\shorttitle{FAINT DWARFS IN OTHER GROUPS}
\shortauthors{SPELLER ET. AL.}

\begin{document}
\title{Faint Dwarfs in Nearby Groups}
\author{Ryan Speller\altaffilmark{1}, James E. Taylor\altaffilmark{1}}

\altaffiltext{1}{Department of Physics and Astronomy, University of Waterloo, Waterloo, ON Canada; rspeller@uwaterloo.ca, taylor@uwaterloo.ca}

\begin{abstract}

The number and distribution of dwarf satellite galaxies remain a critical test of cold dark matter-dominated structure formation on small scales. Until recently, observational information about galaxy formation on these scales has been limited mainly to the Local Group. 
We have searched for faint analogues of Local Group dwarfs around nearby bright galaxies, using a spatial clustering analysis of the photometric catalog of the Sloan Digital Sky Survey (SDSS) Data Release 8. Several other recent searches of SDSS have detected clustered satellite populations down to $\Delta m_r \equiv ({m}_{r,\, {\rm sat}} -\, {m}_{r,\, {\rm main}}) \sim 6$--$8$, using photometric redshifts to reduce background contamination. SDSS photometric redshifts are relatively imprecise, however, for faint and nearby galaxies. Instead we use angular size to select potential nearby dwarfs, and consider only the nearest isolated bright galaxies as primaries. As a result, we are able to detect an excess clustering signal from companions down to $\Delta m_r = 12$, four magnitudes fainter than most recent studies. We detect an over-density of objects at separations $< 400$\,kpc, corresponding to about $4.6 \pm 0.5$ satellites per central galaxy, consistent with the satellite abundance expected from the Local Group given our selection function. Although the sample of satellites detected is incomplete by construction, since it excludes the least and most compact dwarfs, this detection provides a lower bound on the average satellite luminosity function, down to luminosities corresponding to the faintest ``classical'' dwarfs of the Local Group.  

\end{abstract}


\section{Introduction}

One of the great challenges of cosmology is to explain the range of galaxy properties
 observed in the present-day universe. Galaxy formation has long been expected to have a natural cutoff on small scales, as gas cooling becomes inefficient at the low virial temperatures expected in small structures. This argument provided the original framework for understanding why individual galaxies have the range of masses and sizes they do, with fairly well defined upper and lower limits \citep{Silk, ReesOstriker, WhiteRees}. The subsequent development of cosmological models dominated by cold dark matter (CDM) made the argument more precise; the basic framework for CDM galaxy formation, as outlined in \citet{WhiteFrenk}, combines hierarchical structure formation in the dark sector with the physics of gas cooling and star formation in the baryonic sector. In the two decades since this picture was introduced, it has become increasingly clear that the mass function of dark matter halos predicted by CDM models is very different from the galaxy luminosity function observed in the field \citep[e.g.~][]{KWG,Klypin99,Moore99,Benson03,Behroozi12}. In particular, if the universe is full of the small-scale structure predicted in CDM models, most of this structure must remain unilluminated by stars. 

While the existence of a lower cutoff scale for galaxy formation is clear, the detailed nature of the cutoff is much less so. In principle, the cutoff scale could be sharp and well-defined, such that all halos above some mass and none of the halos below that mass are occupied by galaxies. Alternately, it could be that star formation gradually dies out over a broad range of halo mass. In the latter case, we can think of galaxy formation on small scales as being {\it stochastic}, either in the sense that it is truly random, or in the sense that that it is determined by one or more  `hidden'  variables other than mass. There are several plausible candidates for a hidden variable controlling galaxy formation, including merger history, age, environment, or star-formation history. 

The least massive galaxies we know of are dwarf satellites of the Milky Way or the Andromeda galaxy (M31), located within a few hundred kiloparsecs of us. In the CDM model they should correspond to subhalos within the larger dark matter halos of the two main galaxies of the Local Group. The correspondence between galaxies and CDM structure is particularly puzzling on these scales, however. The velocity dispersions of the Local Group satellites suggest that at least some of them may occupy relatively low-mass subhalos, but simulations predict many more subhalos on these scales than there are known satellites \citep{Moore99,Klypin99}. Is this an indication of stochasticity in dwarf galaxy formation? If all the CDM structure predicted by simulations is truly present in the halo of the Milky Way, what sets apart the few subhalos where dwarf satellites have formed? The paucity of observed dwarf satellites has been called the ``missing satellite'' problem, but in a sense it is the opposite. Given the many mechanisms at work to suppress galaxy formation on small scales -- supernova feedback \citep{DekelSilk,Maschenko08,Governato10}, reionization \citep{Efstathiou92,BarkanaLoeb,Bullock00,Gnedin06}, `harassment' \citep{Moore96}, ram-pressure stripping \citep{GunnGott,Nichols11} and/or tidal forces \citep{TB01,Mayer06,Lokas} -- the real question is perhaps, why do we see any dwarf satellites at all?
 
 A major obstacle to resolving the missing satellite problem conclusively is the absence of data for a greater number of groups. Our only reasonably complete samples of brighter satellites are around the two main galaxies of the Local Group, or a handful of other nearby systems \citep{Karachentsev05}. In the case of the extremely low surface-brightness `ultra-faint'  dwarfs with $M_V\lesssim -8$ \citep{Willman05, Belokurov06, Zucker06}, only the Milky Way's population is known, and even it is likely to be significantly incomplete \citep{Koposov08,Tollerud08}. This raises the question of how representative the Milky Way or Local Group satellite populations are. 

In particular, there has been much recent discussion of the relative frequency of bright satellites such as the LMC, SMC, M33, or M32, and what this implies about the Local Group. An excess of bright satellites relative to model predictions was noted even in early semi-analytic models of the Local Group \citep[e.g.][]{Benson02}. More recently, numerical simulations indicated that galaxies like the Milky Way should only very rarely host a pair of satellites as massive or bright as the Magellanic Clouds (\citealt{BK10, Busha11}; see also \citealt{Gonzalez}). It has now been confirmed by several observational studies that the Milky Way is indeed unusual in this respect (\citealt{JamesIvory11,Liu2011,Robotham12} -- although see also \citealt{Tollerud11}). The two Magellanic Clouds are also on similar orbits and probably fell in together; they may have once formed a group including other Milky Way satellites \citep{LyndenBell,Nichols11b,Sales11}. Very recently, a similar orbital grouping has been discovered around M31 \citep{Ibata2013}. In short, both of the two main Local Group satellite systems appear to have their idiosyncrasies. The models invoked to explain the abundance of satellite galaxies, however, normally assume the Local Group populations are typical, so this an important assumption to verify. 

Beyond average numbers, one might also wonder what, if any, is the relationship between the properties of the satellite population and the properties of the primary galaxy they orbit, e.g.~its total stellar mass, morphology, or recent merger history. To address this issue it is important to find Local Group analogues, isolated bright galaxies with populations of dwarf galaxies that are well-enough sampled to determine their clustering scale length, the amplitude of the satellite luminosity function, and the distribution of dwarf colours, morphology and other properties, information we have so far only for the Local Group dwarfs.

There has been much recent progress identifying satellites around normal galaxies. The easiest to study are bright companions like the Magellanic Clouds, M33 or M32, which are only 3--4 magnitudes fainter than their primaries (e.g.~\citealt{Liu2011,Prescott11,Sales2012,Guo2012} -- see \citealt{Wang12} for earlier references). Fainter satellites have also been detected statistically using large-area ground-based surveys, as an excess of faint objects clustered around a population of brighter primary galaxies \citep{Carlberg,Lares2011,Guo2011,Wang12,Jiang, Strigari2012}. Finally, space-based imaging has allowed these searches to be extended to higher redshift. In recent work, for instance, \cite{Nierenberg2011,Nierenberg2012} have detected satellites extremely close to central galaxies, by selecting smooth early-type galaxies as primaries and subtracting  from the image a model fit to their light.
 
So far, most of these studies have considered large but relatively distant samples of primaries, selected in all but a few cases from the Sloan Digital Sky Survey \citep[SDSS --][]{SDSS}. As a consequence, measurements of the satellite luminosity function have generally been limited to satellites 6--8 magnitudes fainter than their primary (corresponding to $r$-band magnitudes $M_r = -14$ to $-12$). The missing satellite problem becomes most severe further down the luminosity function, however, so it is important to seek out fainter satellites in a large sample of nearby systems. In particular, searches for companions should target primary galaxies at distances intermediate between those of the few nearest groups (5\,Mpc or less) and those typical of SDSS samples (200--400\,Mpc or more, e.g.~\citealt{Lares2011,Guo2011, Strigari2012}).

In this paper, we search for excess clustering of SDSS galaxies around a sample of primary galaxies within 42\,Mpc of us. At these very small distances, we can use size cuts to eliminate many of the more distant background galaxies, making the selection of local dwarf candidates relatively efficient. The proximity of these galaxies also allows us to probe the satellite luminosity function 4--5 magnitudes below most previous results. The outline of the paper is as follows. In section 2 we describe our primary and satellite samples; in section 3 we measure the clustering signal and explore its dependence on primary luminosity and morphology, and in section 4 we determine the relative luminosity function per primary. In section 5 we summarize our results and discuss the prospects for future searches using data from deeper surveys.

\section{The Samples}

Our goal is to find faint satellites around systems roughly comparable to the primaries of the Local Group, and thus we want to search around nearby primaries with luminosities and/or stellar masses comparable to the Milky Way and M31. Furthermore, the primary sample must overlap with a large-area survey such as the SDSS, and contain enough galaxies to obtain reasonable statistics. A recent catalogue that satisfies these requirements is the parent catalogue of the Atlas-3D survey \citep{Atlas3D}. We discuss this primary sample, and our construction of a background/satellite sample, below.

\subsection{Primaries}
Our sample of primary galaxies is drawn from the Atlas3D parent catalogue of \cite{Atlas3D}. The Atlas3D survey targeted nearby, bright early-type galaxies. These were selected from a larger parent sample of bright elliptical, S0, and spiral galaxies that was designed to be volume-limited above a certain stellar mass. The parent sample was constructed using $K_s$-band magnitudes from the 2-Micron All-Sky Survey \citep[2MASS;][]{2MASS}, and the best distance estimates available from the literature. Given these distance estimates, the parent catalogue is complete down to $M_{K_{s}} = -21.5$ out to a distance of 42\,Mpc, corresponding to a stellar mass limit of  $\textrm{M}_{\star} \geq 6 \times 10^9 \textrm{M}_{\odot}$. It covers 37\%\ of the sky, corresponding to a volume of $1.16\times 10^5$\,Mpc$^3$ 
out to 42\,Mpc. The full catalogue contains 871 galaxies, including 68 ellipticals, 192 S0s, and 611 spirals

Many of the galaxies in the parent sample are members of nearby galaxy clusters such as Virgo. In order to restrict our search to systems more analogous to the Local Group, we apply an isolation criterion to the parent sample. First, we remove M31 from the sample, since it is so nearby that its projected virial radius overlaps with many background systems.  We then make a series of isolation and quality control cuts on the sample: 
\begin{enumerate}
\item{We remove any member of the sample that is within a (3-D) distance of 1.5\,Mpc of another member. This reduces the catalogue to 356 galaxies, and excludes most cluster members.}
\item{We remove objects that are not in the SDSS footprint, or are in badly masked regions or regions of incomplete coverage. This reduces the sample to 282 galaxies.}
\item{We remove a further 5 galaxies which are within 5$^{\circ}$ of the centre of the Virgo cluster in projection, and 3 galaxies which are within 3$^{\circ}$ of the centre of the Coma or Leo clusters.}
\end{enumerate}
After making these cuts, we are left with 274 primaries, all isolated massive galaxies within 42\,Mpc with good coverage in SDSS. 

The Atlas3D catalogue provides distances, morphological T-types, and (2MASS) $K_s$ magnitudes for the sample. We also obtain total $r$-band magnitudes from SDSS where possible. In cases where the latter are poorly determined due to the size or brightness of the galaxies, we estimate the total $r$ magnitude from the RC3 $B$ magnitude \citep{RC3} using the colour conversions and mean colours for different morphological types given in \cite{Fukugita}, or we use the 2MASS $K_s$ magnitude and assume the $r-K_s$ colour is equal to the mean of the entire sample, $r-K_s = 2.9$. The dispersion in $r-K_s$ colour for the sample is $\pm 0.9$, so this gives an indication of the possible uncertainty in the final $r$ magnitude for these systems.

The primary $K_s$ magnitudes range from $-21.5$ to $-25.6$, with an average value $\langle M_K\rangle = -22.88$. Given that the absolute magnitude of the Sun in the $K_s$ band is estimated to be $M_{K,\odot} = 3.29$ \citep{Blanton}, the primary luminosities range from $8.2\times 10^9$ to $3.6\times 10^{11} L_{K,\odot}$, with an average of $2.9\times 10^{10} L_{K,\odot}$. \cite{WIlliams} derive a mean $K_s$-band mass-to-light ratio of $(M /L)_{K_s} = 1.09$ for a sample of 14 S and S0 galaxies, with a rms scatter of  30\%. The appropriate value for ellipticals may be slightly higher, although it shows little systematic variation with colour in the sample of \cite{WIlliams}. We conclude that our primaries have a mean stellar mass of $\sim 3$--$3.5\times10^{10} M_\odot$, with a range of $1\times 10^{10} M_\odot$ to $4\times 10^{11} M_\odot$. Using the stellar-to-halo-mass ratio determined by \cite{Leauthaud} from a combination of galaxy-galaxy lensing, clustering and abundance matching, this should correspond to an average halo mass of $1\times 10^{12} M_\odot$, and a range from $4\times 10^{11} M_\odot$ to $1\times 10^{14} M_\odot$ or more. The abundance matching model of \cite{Guo_abundance} also predicts similar values. We note that our isolation criteria will remove galaxies in the most massive systems from the primary sample; as a result we expect our most luminous systems to occupy halos with masses of a few times $10^{13} M_\odot$ or less. The corresponding virial radii for our sample in a concordance $\Lambda$-CDM cosmology range from 200--600\,kpc, or 0.39$^\circ$--1.15$^\circ$ at 29.7\,Mpc, the mean distance to our primaries.

\subsection{Satellites}

Around our final sample of primaries, we search for potential satellites at projected separations up to 1\,Mpc (corresponding to $2.86^{\circ}$ at a distance of 20\,Mpc, or $1.43^{\circ}$ at 40\,Mpc). In principle, we could apply a cut on photometric redshift to keep only those objects likely to be close to our primary sample, which extends out to $z \sim 0.01$. In practice, SDSS photometric redshifts have large uncertainties over the redshift and magnitude ranges of interest;  after examining the photometric redshift estimates of objects at known distances, we choose to make a relatively conservative cut at $z_{\rm phot} = 0.15$. We use an SQL query of the SDSS Catalogue Archive Server\footnote{CAS -- \url{http://skyservice.pha.jhu.edu/casjobs}} to retrieve any object that is within 1\,Mpc projected separation from a primary galaxy and has a photometric redshift of 0.15 or less. 

Regions of SDSS DR8 have been masked due to halos from bright stars or other artifacts. In order to obtain reliable background statistics, such as galaxy counts in circular annuli, it is necessary to identify the boundaries of masked regions and correct for the section of each annulus that has been cut out of the data. To detect masked regions automatically, we construct a square grid covering each field and search for cells devoid of catalogue objects. The cell size is chosen so that each cell should contain 4 objects on average, in the case of a uniform distribution. Masked sections are identified as sets of two or more adjacent cells that contain no objects. Once a masked section is detected, any adjacent empty cells are added to it to find the total extent of the masked region. 

For objects in the vicinity of each primary, we record magnitude and size. SDSS provides several measures of galaxies' apparent magnitudes. Of these, we use composite model (\textit{cmodel}) magnitudes. These are calculated from exponential and de Vaucouleurs fits in each photometric band, using the linear combination of the two that best fits the image. \cite{Strauss02} report that \textit{cmodel} magnitudes are less dependent on local seeing variations than other measures, and are thus a good proxy to use as a universal magnitude for all types of objects. These magnitudes are extinction corrected; we do not K-correct them since the sample is local. For size, we use the exponential scale radii provided by SDSS to quantify galaxy size. 

In order to reduce the background of distant galaxies included in the satellite sample, the SDSS data are cut by magnitude, colour and size: 
\begin{enumerate}
\item{} First, since the limiting $r$-band photometric magnitude in SDSS is about 22, galaxies that appear to be dimmer than this, or brighter than an apparent magnitude of $m_r = 10$, are eliminated as well. 
\item{}Second, we apply a colour cut, removing any objects with extreme colours and restricting our sample to galaxies with $$ -1 \le (g-r) \le 0.85\,.$$ This corresponds broadly to the colour range of local galaxies, allowing for large errors in colour at faint magnitudes.
\item{}The most important cut on the data is a magnitude-dependent size cut. Only galaxies with exponential scale lengths $r_{\rm exp}$ and magnitudes $m_r$ such that:
\begin{equation}
24 - {{r_{exp}}\over{1\arcsec}} < m_r < 30 -  {{r_{exp}}\over{1\arcsec}}
\end{equation}
are included in the sample, while larger and smaller objects are excluded. This cut has the effect of removing 98\% of the remaining background.
\end{enumerate}

Local Group dwarfs are assumed to be indicative of the satellites we expect to find around our primaries. To test our cuts, we define a `Local Group comparison sample'. This consists of all the known members of the Local Group that would pass our cuts in size, magnitude, and colour, if they were observed from a distance of 25\,Mpc away. (We note that the inventory of Local Group members is likely to be incomplete at the faint end, at large galactocentric distances, and/or at low latitudes \citep{Koposov08,Tollerud08,McConnachierev}, but these problems affect mainly the faintest objects, with $M_V \lesssim -10$.)  Numbers in the comparison sample are divided by two, since there are two bright primaries contributing to the Local Group population. We refine our cuts using this sample; in particular the parameters in our size-magnitude cut (equation 1) were determined iteratively, so as to maximize the contrast between the SDSS field population and the Local Group comparison sample. 

\section{The Clustering Signal}

We search for a satellite population by measuring the clustering of objects in the background catalogue with respect to our primaries. Our method is similar to that in other recent studies \citep[e.g.][]{Liu2011,Lares2011,Guo2011}. Possible systematics have been considered in detail by \cite{Chen2006} and \cite{Wang12}.

\subsection{Results for the Full Sample}
After the background catalogue is filtered using the cuts described above, we calculate projected separations ${\rm R_p}$ from the central primaries in linear units at the distance of the primary. Galaxies are binned by separation from the primary in bins of width 50\,kpc, and the area on the sky in an annulus corresponding to each separation bin is calculated. We correct these areas for masking, as described above. In each bin, we calculate the surface density of galaxies $\Sigma$, and the density contrast relative to the background density ${\overline{\Sigma}}$, $\delta_\Sigma \equiv (\Sigma/{\overline{\Sigma}}) - 1$. We estimate the background density using a set of outer annuli with projected separations ${\rm R_p} =$ 0.5--1\,Mpc. We note that the innermost annulus (${\rm R_p} =$ 0--50\,kpc) may be contaminated by bright globular clusters or HII regions associated with the primary, and faint objects close to bright, extended ones can also have systematic errors in their SDSS photometry \citep{Wang12}; thus we will exclude this bin when calculating the cumulative excess within a given projected radius.

\begin{figure*}
\plottwo{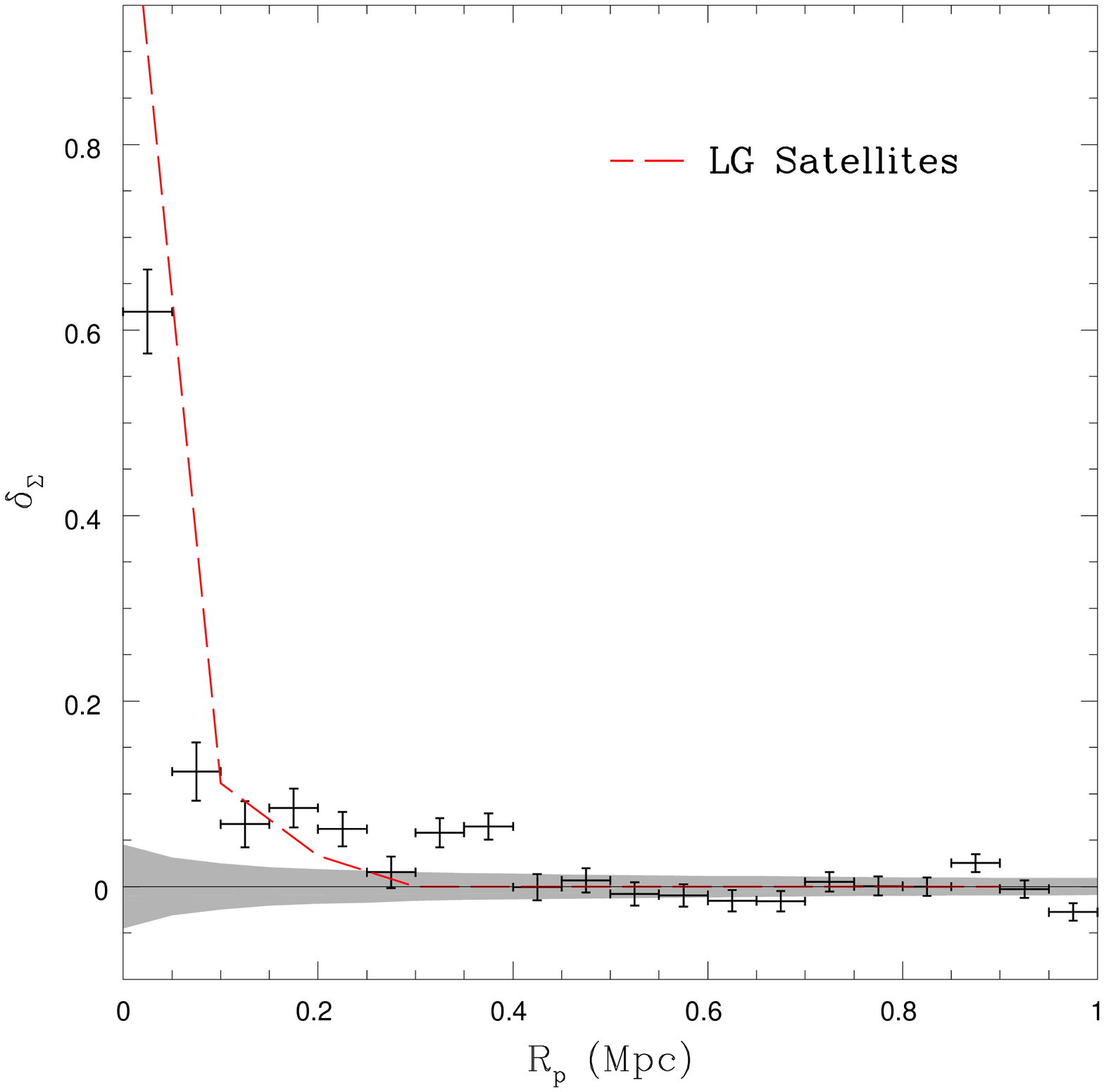}{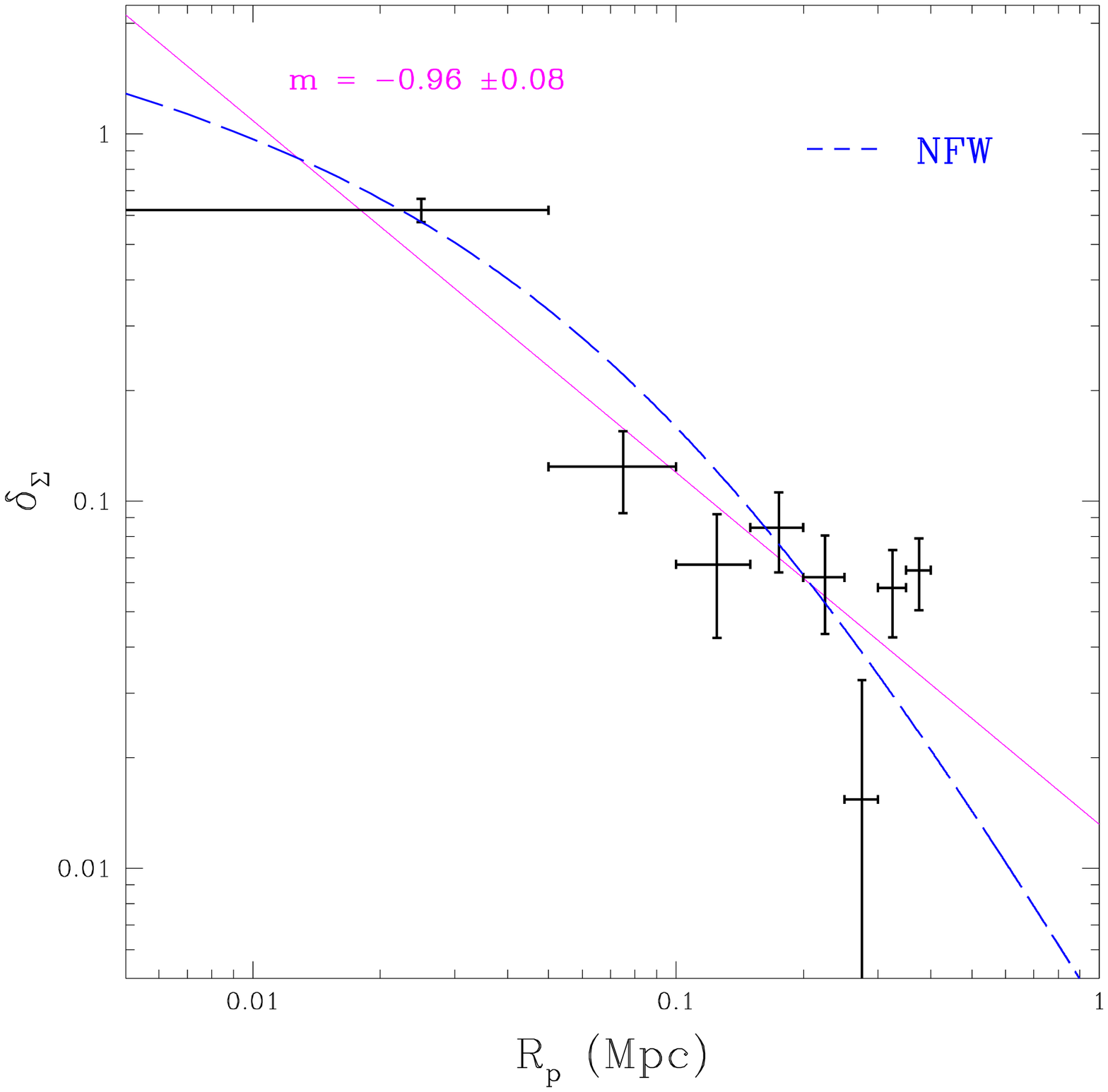}
\caption{{\it Left:} The surface density contrast profile for the full sample. Grey shading indicates the rms scatter in the background surface density. The dashed line indicates the surface density of the Local Group comparison sample (see text), averaging over 3 projections. Horizontal error bars indicate bin widths, while vertical error bars indicate the uncertainty in $\delta_\Sigma$, assuming Poisson errors in the galaxy counts. {\it Right:} The inner bins on a logarithmic scale, to highlight the radial dependence of the density excess. The solid line shows a linear fit to the inner bins of slope $m = -0.98 \pm 0.08$; the dashed curve shows a projected NFW profile of concentration $c = 10$. 
\label{fig01}}
\end{figure*}

Figure \ref{fig01} shows the surface density contrast as a function of projected separation. A positive contrast, corresponding to an over-density of objects at projected separations ${\rm R_p} \le 400$\,kpc, is detected at S/N $\sim9$. The rms scatter in the background surface density is shown as the grey shading in the left-hand panel. We also plot the corresponding surface density contrast for the Local Group comparison sample (dashed red line in the left-hand panel). We see the detected excess is comparable to that expected for a system like the Local Group, exceeding it slightly at large radii (200--400\,kpc). (Some of the contribution from large radii may come from systems more massive than the MW or M31, as dicussed in section 3.2 below.)

In the right-hand panel of Fig.~\ref{fig01}, we plot the projected density contrast on a logarithmic scale, to emphasize the radial dependence. The mean logarithmic slope is $m \equiv {\rm d}\ln \delta_\Sigma /{\rm d}\ln r = -0.98 \pm 0.08$, consistent with a projected isothermal profile ($m = -1$), but also with a projected NFW profile (dashed blue curve). This is roughly as expected if satellites trace the mass of the halo surrounding each primary, although previous work indicates that the radial profile will depend on the details of the primary and satellite samples (\citealt{Lorrimer, Chen2006,Chen08,Jackson,Sales11, Lares2011,Tal, Nierenberg2011,Jiang} -- see \citealt{Guo2012} and \citealt{Nierenberg2012} for recent discussions).

The average excess satellite count per primary in the radial bin $b$, $\Delta N^b$, is determined as: 
\begin{equation}
\Delta N^b = \frac{1}{{N_{gal}}}\left(N^b_{inner} - \frac{A^b_{inner}}{A_{outer}} N_{outer}\right)
\label{equ:num}
\end{equation}
Where $N_{gal}$ is the total number of primaries, $N^b_{inner}$ is the number of galaxies in inner bin $b$, $A^b_{inner}$ is the annulus area in bin $b$ in the inner region, $A_{outer}$ is the total area of the outer region, and $N_{outer}$ is the number of galaxies in the outer region. The inner bins range from 50--500 kpc in projected radius around each primary, while the outer region is the annulus ranging from 0.5 to 1 Mpc in each case. The error in the final count per primary is calculated assuming Poisson errors on $N^b_{inner}$  and $N_{outer}$ and using the usual rules for error propagation; we ignore any contribution from uncertainties in the areas $A^b_{inner}$ and $A_{outer}$ which might arise from our masking corrections, as these are typically negligible. 

The background correction $-({A^b_{inner}}/{A_{outer}}) N_{outer}$ represents the mean surface density of the outer regions ($R_{\rm p} = 0.5$--1 Mpc). Subtracting this term should remove both the contribution from uncorrelated galaxies and the contribution from large-scale (``two-halo") clustering, which dominates on scales of a few Mpc \citep[e.g.][]{Liu2011}. There has been some discussion of whether using a locally determined background in clustering measurements introduces bias and/or reduces the signal-to-noise \citep{Wang12}. We have tested the effect of determining the values of $A_{outer}$ and $N_{outer}$ individually for each primary, or calculating an average background for all primaries first, and then scaling and subtracting this following Eqn.~\ref{equ:num}. We find that the clustering signal is very similar and the differences minor in both cases. The signal-to-noise is marginally higher using an average background, so we have used this method in what follows. We also note that the inner edge of our background region may overlap slightly ($\sim 15\%$ in projected area) with the virial volume for the few most massive primaries in our sample, so the excess counts may be slightly underestimated for these objects. 

\begin{figure}
\plotone{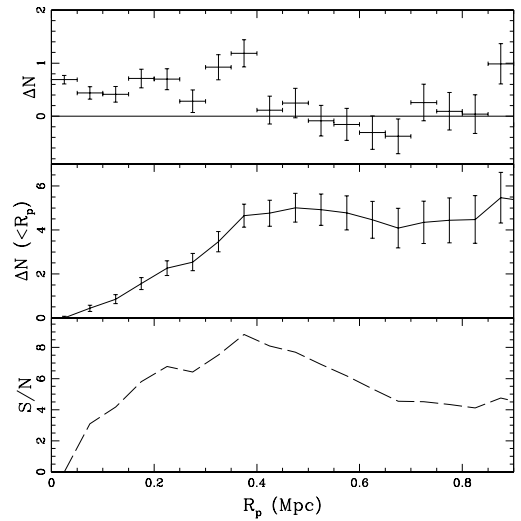}
\caption{Average excess counts per galaxy in bins of projected radial separation ${\rm R_p}$, for the full sample of 274 primaries. The top panel shows the excess counts in each bin, the middle panel shows the cumulative excess counts within a given projected separation, and the bottom panel shows the signal-to-noise ratio of the cumulative excess.\label{fig02}}
\end{figure}

Figure \ref{fig02} shows the excess counts per bin (top panel), the cumulative excess within a given radius (middle panel), and the signal-to-noise ratio of the cumulative detection (bottom panel). Errors on the cumulative counts are simply the errors on individual bins, added in quadrature. The signal-to-noise ratio we define as the cumulative excess within some radius, divided by its error. In calculating the cumulative counts and signal-to-noise, we exclude the innermost bin (${\rm R_p}=\,$0--50\,kpc) to avoid contamination from the primary, as discussed previously.
Integrating out to ${\rm R_p} = 400$ kpc where we reach the best signal-to-noise, we find a net excess of $4.65 \pm 0.53$ satellites per central galaxy, at S/N = 8.8. 

\subsection{Dependence on Primary Luminosity}

\begin{figure}
\epsscale{1.00}
\plotone{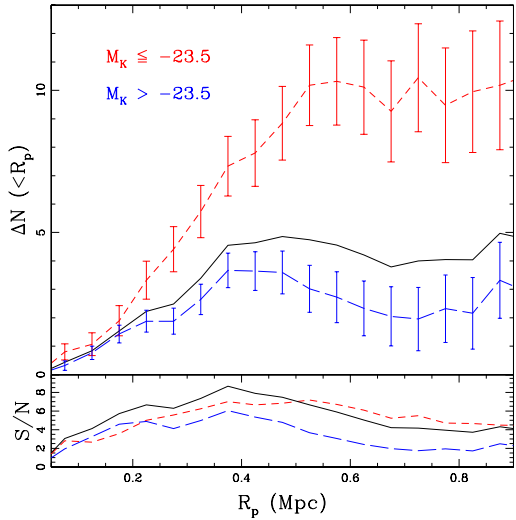}
\caption{Cumulative excess counts per galaxy in bins of projected radial separation, as a function of primary magnitude. The top panel shows the cumulative excess, and the bottom panel shows the corresponding signal-to-noise ratio. Red short-dashed lines indicate bright primaries ($M_K \le -23.5$),  blue long-dashed lines indicate faint primaries ($M_K > -23.5$), and black solid lines indicate the counts for the whole sample. \label{fig03}}
\end{figure}

To test the dependence of the satellite population on primary luminosity, we split the full sample into two subsamples: 66 bright primaries with $M_K \le -23.5$ and 208 fainter primaries with $M_K > -23.5$. The average $K_s$ magnitudes for the two samples are $-24.1$ and $-22.5$, corresponding to luminosities of $9.3\times 10^{10}$ and $2.1\times 10^{10} L_{\odot, K}$ respectively. The top panel of Figure \ref{fig03} shows the cumulative number over-density profiles for each subsample. Red (short-dashed) lines indicate the brightest primaries, blue (long-dashed) lines indicate fainter primaries, and black (solid) lines show the average for the whole sample. The bottom panel shows the S/N of each cumulative detection. As before, our error bars assume Poisson errors in the counts per bin. We find a strong dependence in the number of associated satellites on primary magnitude, with the brightest 25\% of the sample having a cumulative count of associated satellites 
3 times higher than the remaining 75\% of the sample. The shape of the cumulative number profile also changes slightly with primary magnitude; in bright systems counts appear to rise out to projected separations of 500--600\,kpc, whereas for the fainter primaries they reach a maximum by 400\,kpc. Primaries with $K_s$-band magnitudes $M_{K_{s}} \leq -23.5 $, shown in red, have the greatest signal of $10.2 \pm 1.4$ galaxies in excess of the background at a signal-to-noise of $\sim 7$ in the 500--550\,kpc projected radial separation bin. Primaries with $M_{K_{s}} > -23.5$ have $3.7 \pm 0.6$ satellites each within a projected radius of 350--400\,kpc, where the S/N reaches a maximum of $\sim 6$. 

\subsection{Dependence on Primary Morphology}

\begin{figure*}
\epsscale{1.00}
\plottwo{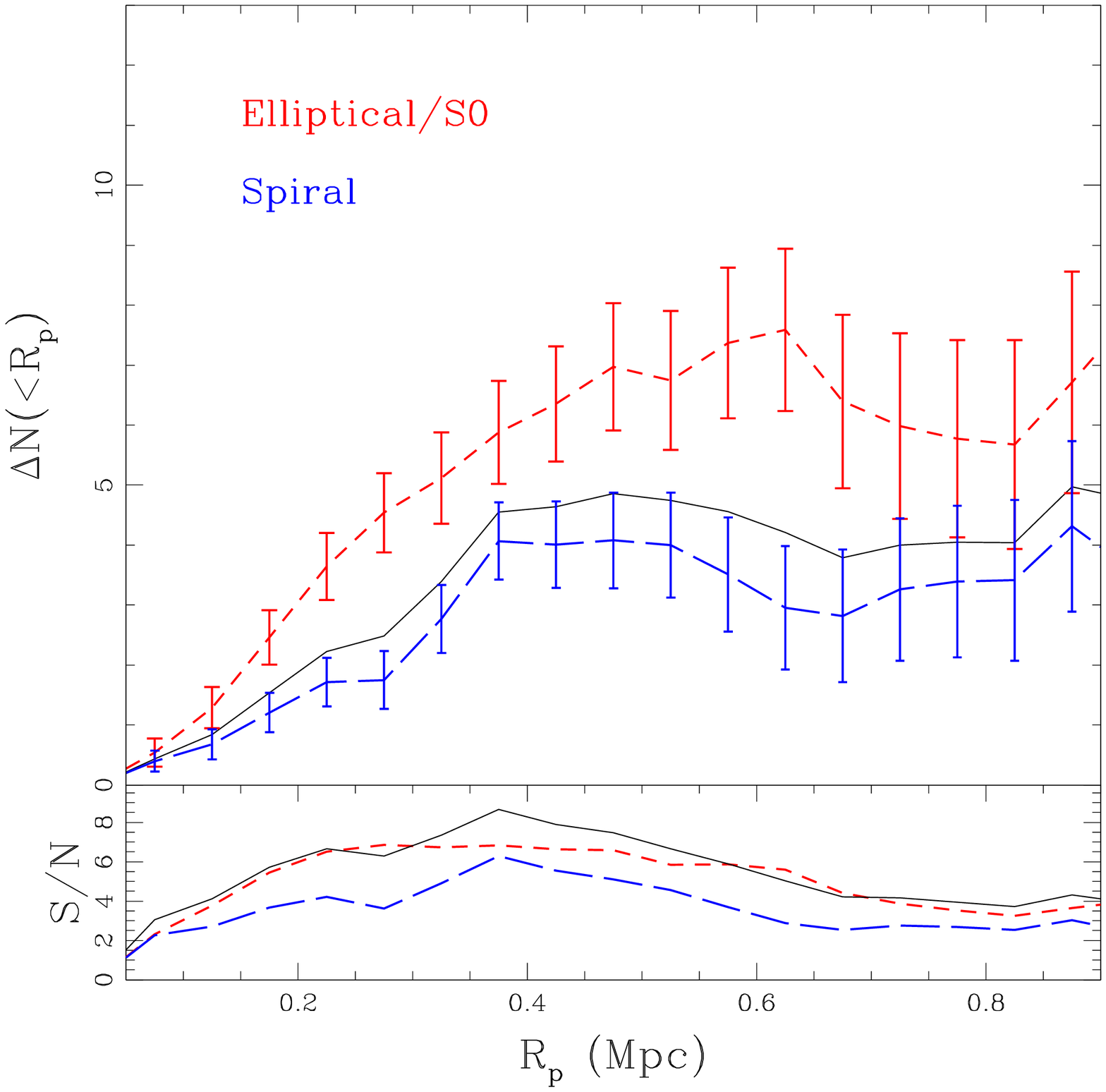}{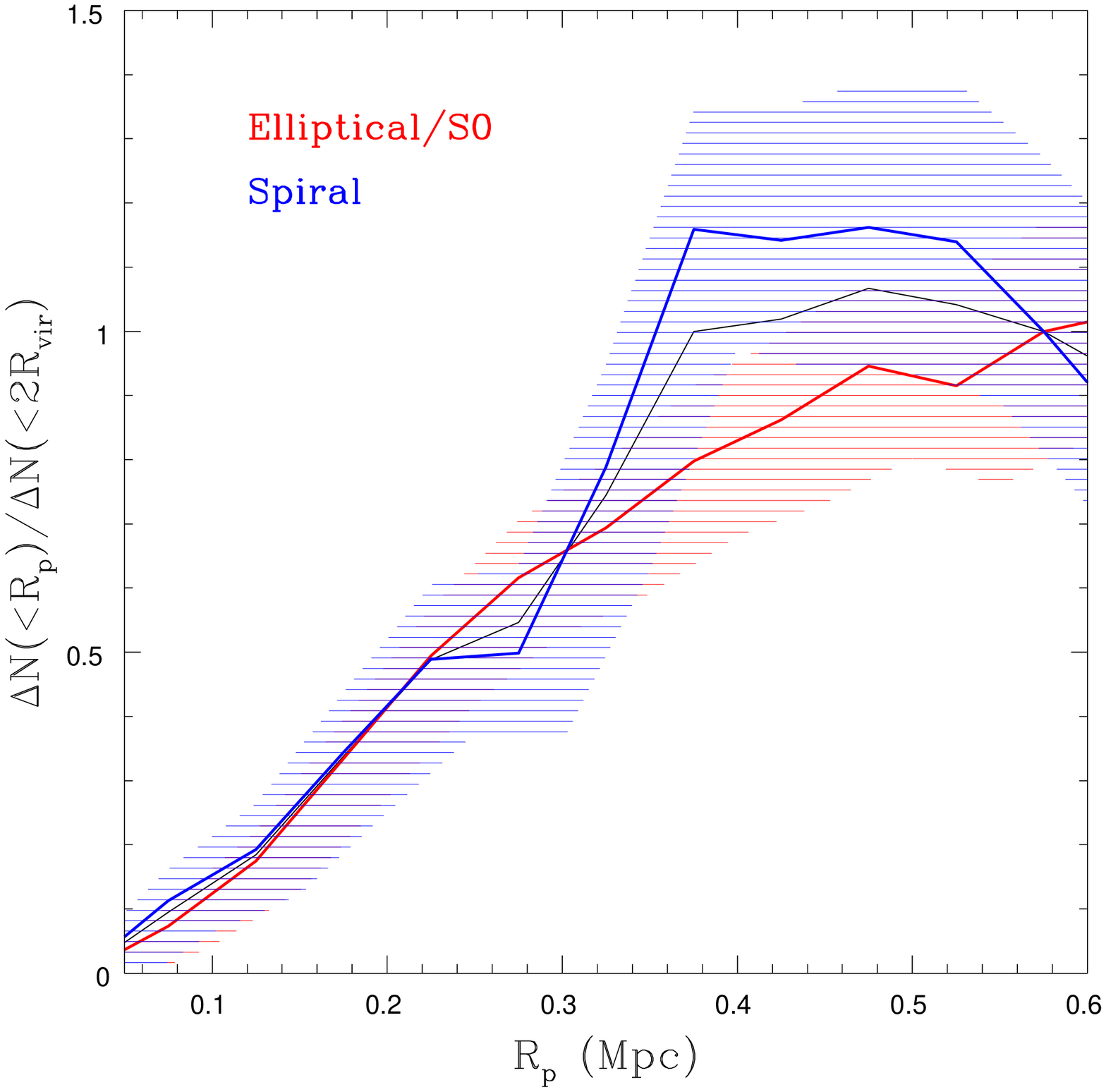}
\caption{{\it Left}: As figure \ref{fig03}, but for primaries split by morphology. Red short-dashed lines indicate ellipticals and S0s ($T \le 0$),  blue long-dashed lines indicate spirals ($T > 0$), and black solid lines show the counts for the whole sample. 
{\it Right:} Cumulative excess counts for the two morphological types. Note these are normalized at 550--600\,kpc (approximately $2R_{vir}$ for the median halo mass). \label{fig04}}
\end{figure*}

We also split our primary sample by morphology into elliptical/S0 ($T \le 0$; 73 in total) and spiral ($T > 0$; 201 in total) types, as determined by the Atlas3D survey \citep{Atlas3D}. These two samples have almost identical mean magnitudes and rms scatter in the K-band ($M_K = -22.87  \pm  0.88$ and $M_K = -22.88  \pm  0.90$ respectively). As before, the cumulative excess counts for each case are shown in Figure \ref{fig04}. Elliptical primaries appear to have more satellites than spirals ($\sim7.6 \pm 1.4$, vs.~$4.1 \pm 0.65$), spread out over a slightly larger spatial scale ($\sim$650\,kpc vs.~400\,kpc, or a factor of 1.6). Although the contrast between the two subsamples is less marked than in the previous section, the difference in abundance and clustering scale suggests that they occupy halos of different mean mass. The right-hand panel of figure \ref{fig04} shows the radial distributions relative to their value at 600\,kpc (corresponding to approximately 2\,$R_{vir}$ for our sample on average), to emphasize the steeper rise in counts with radius around spiral primaries. Although the signal-to-noise is low in our split samples, if we take the difference in clustering scale at face value it suggests a difference in halo mass of $\sim$4 $( = 1.6^3)$ at {\it fixed stellar mass}. Thus, we find tentative evidence for morphological dependence in the ratio of luminosity to dark matter halo mass. This could indicate differences in the mass-to-light ratios of the stellar populations, or a more fundamental difference in the stellar-to-halo-mass ratio \citep{Leauthaud}. The result is only of marginal ($2.5 \sigma$) significance, however, and it could also be that a few massive systems are biasing our subsample of ellipticals and S0s. With an all-sky sample or deeper imaging around our current sample, we should be able to test for this effect more convincingly.

\subsection{Dependence on Satellite Colour}

Finally, we can attempt to determine how the clustering signal varies with satellite colour. We might expect red and blue satellites to cluster on different scales, given the morphology-density relation observed in the Local Group \citep[e.g.][]{McConnachierev}. In large samples, the clustering amplitude of brighter satellites also depends on secondary as well as primary colour \citep[e.g.][]{Wang12}. Testing colour cuts on our background sample, however, we find that most ($\sim$85\%) of the clustering signal comes from blue ($g - r \le 0.6$) satellites. This is presumably due to our size cut, blue galaxies being larger at fixed luminosity. As it is, the clustering signal for blue satellites is similar to the signal for the whole sample, while the signal-to-noise for red satellites alone is too low to detect any obvious differences.  

\section{The Relative Luminosity Function}

\begin{figure}
\plotone{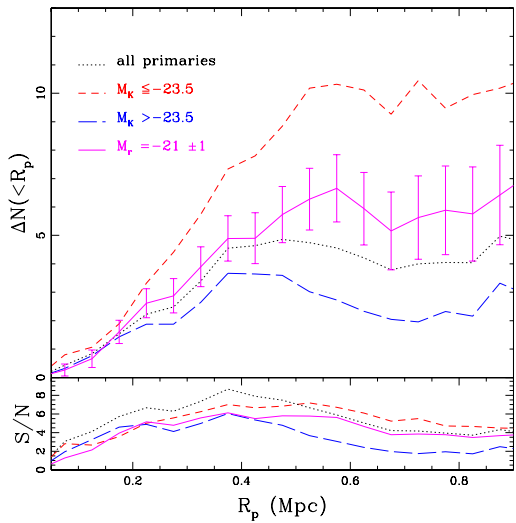}
\caption{{\it Top panel:} Cumulative excess counts per galaxy in bins of radial separation, for primaries with $\rm M_r = -21 \pm 1$ (magenta solid lines with error bars), compared with the bright subsample (short-dashed red lines), the faint subsample (long-dash blue lines), and the entire sample (dotted black lines). {\it Bottom panel:} The corresponding signal-to-noise ratios.\label{fig06}}
\end{figure}

Our final goal is to construct a relative luminosity function for our satellite population, that is the mean number of satellites per primary with magnitudes in some range relative to the primary, $N(\Delta m)$ where $\Delta m = m_{{\rm sat}} - m_{{\rm main}}$. This function, considered previously by many other groups (e.g.~\citealt{Liu2011,Lares2011,Guo2011,Nierenberg2011,Guo2012,Strigari2012} -- see \citealt{Nierenberg2012} for a summary) provides an interesting point of comparison to the relative mass function $N(M_{\rm sat}/M_{\rm main})$ often studied in CDM simulations. In order to compare directly with the previous results we calculate the magnitude difference in the SDSS $r$-band, using primary $r$ magnitudes estimated as discussed in section 2.1.

The Milky Way is the obvious point of comparison for relative luminosity functions, and most previous authors have defined subsamples of their data thought to match its properties, in particular its luminosity in the $r$-band. The luminosity usually assumed corresponds to a total magnitude of $M_{r,MW}\sim -21$, based on the $V$-band magnitude estimate from \cite{vandenBergh} \citep[e.g.][]{Liu2011}. This is comparable to the mean $r$ magnitude of our bright ($M_K \le -23.5$) sample, $\langle M_r\rangle = -21$. The mean halo mass of the bright sample, when estimated as described in section 2.1, is probably 3--4 $\times 10^{12} M_\odot$, however -- that is twice the mass of the Milky Way's halo. Furthermore, our bright sample also includes at least a few systems with much larger halo masses of a few times $10^{13} M_\odot$. Thus, the Milky Way may be intermediate between the average properties of our bright and faint subsamples. To provide a reasonable point of comparison, we define an intermediate sample in the $r$-band, of the 143 primaries with $M_r = -21 \pm 1$. This intermediate sample has a $L_K = 4.7\times 10^{10} L_{\odot}$, suggesting a mean halo mass of 1.5--2$\times 10^{12} M_\odot$, closer to the estimates of the Milky Way's halo mass.

Figure \ref{fig06} shows the excess counts for the `Milky-Way-like' sample with $M_r = -21 \pm 1$ (solid magenta line with error bars), compared with our full sample (dotted black line), and the $K_s$-band bright and faint subsamples (red and blue short and long-dashed lines respectively). The counts are intermediate between the bright and faint samples, and increase out to at least 500\,kpc. Overall, the largest signal-to-noise ratio of $\sim6$ occurs at ${\rm R_p}\sim400$\,kpc, where we find an excess of $4.9 \pm 0.8$ galaxies above the background, (although the signal-to-noise ratio remains similar out to 600\,kpc, where the excess is $6.7 \pm 1.2$ galaxies). For the calculation of the relative luminosity function we will include counts between projected radii of 50 and 400\,kpc, to maximize the signal-to-noise for the Milky-Way-like sample while avoiding possible contamination from
the primary.

\begin{figure*}
\centering
\begin{tabular}{cc}
\includegraphics[height=0.3\textheight]{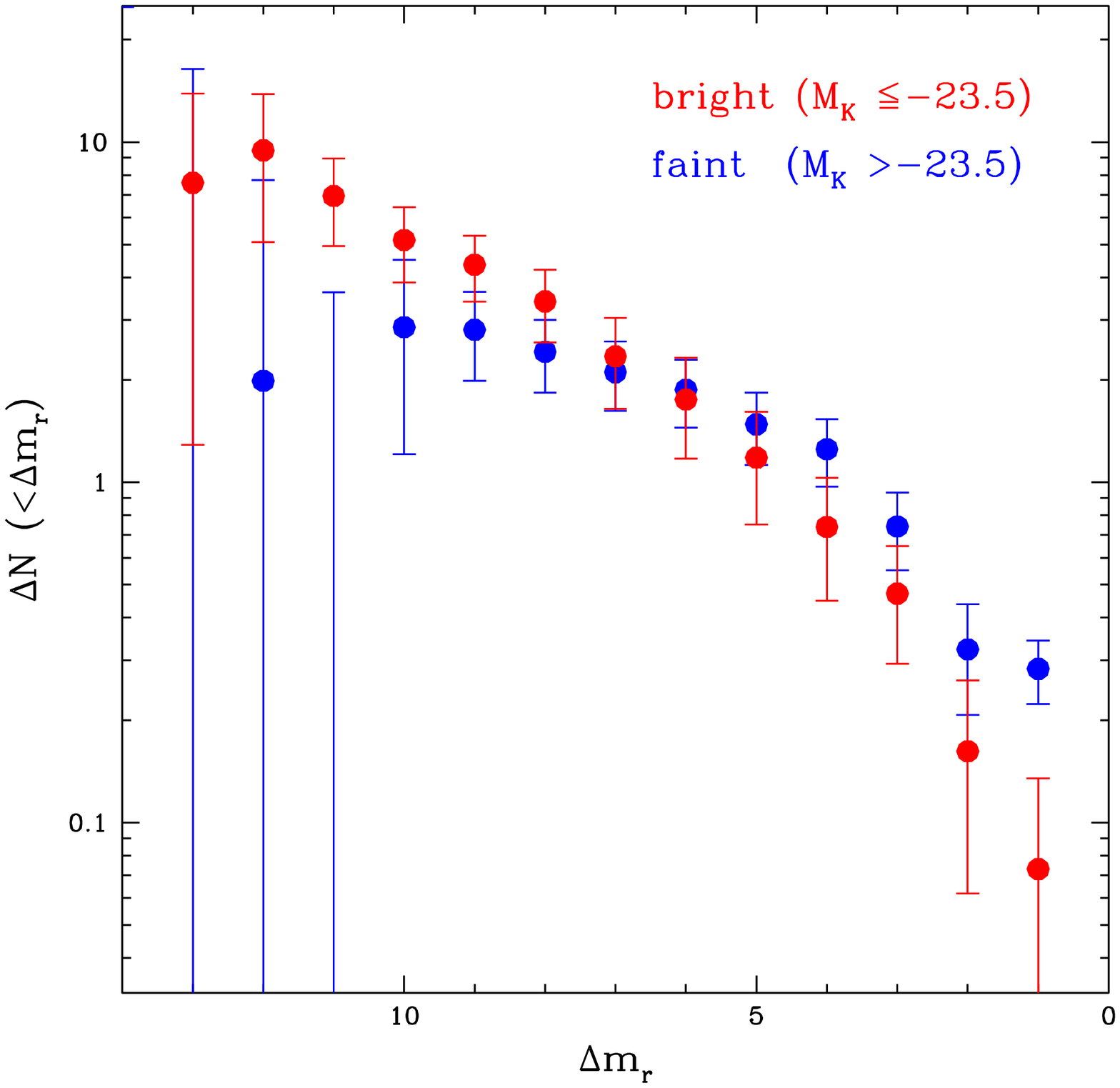} &
\includegraphics[height = 0.3\textheight]{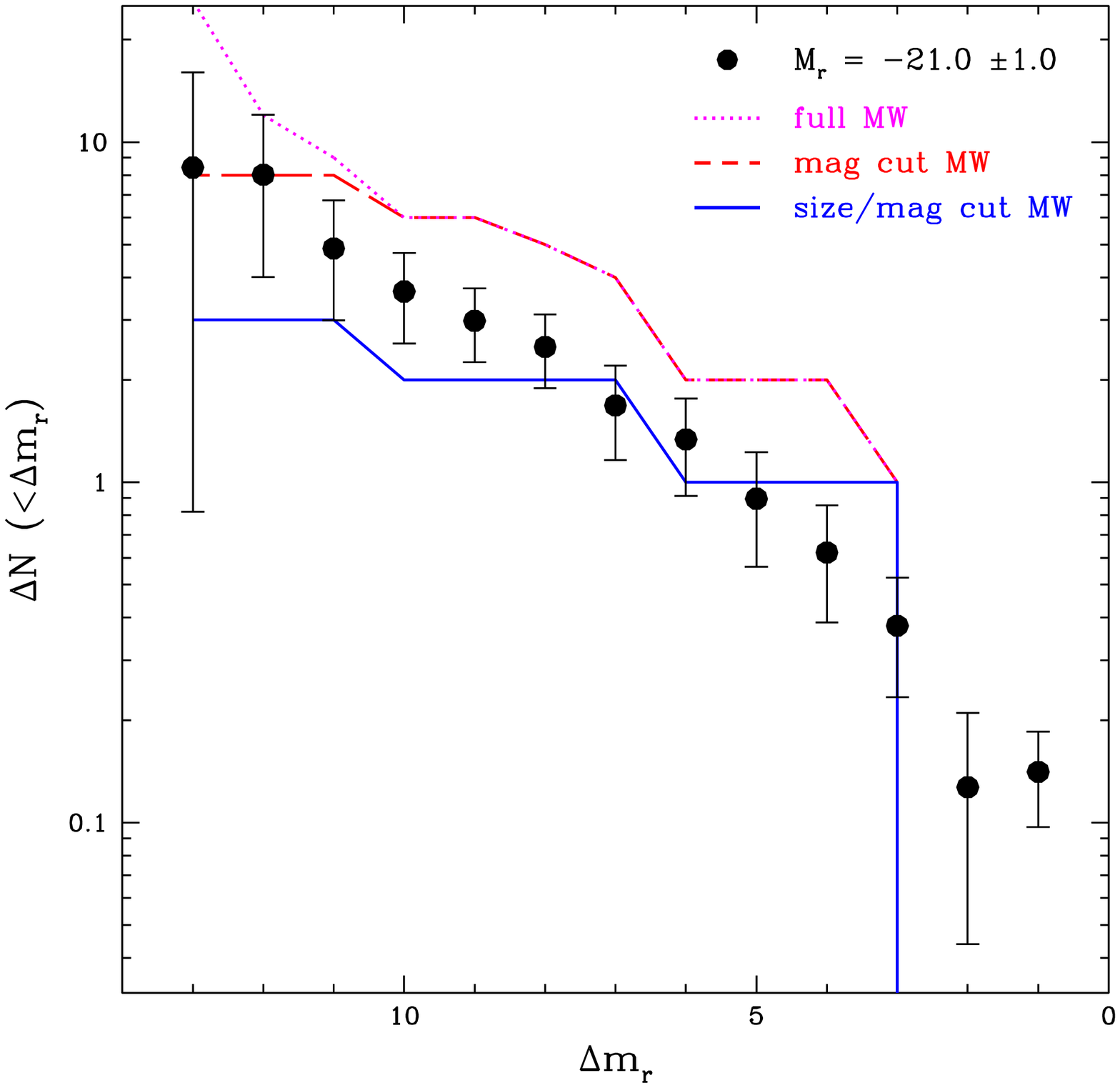}
\end{tabular}
\caption{{\it Left:} The relative luminosity function, that is the excess counts per primary binned by $\Delta m_r$, the difference in $r$ magnitude between the satellite and the primary, and cumulated in magnitude. Red indicates bright primaries; blue indicates faint primaries. Averages and error bars are calculated using simulated background counts and inverse-variance weighting, as described in the text. {\it Right:} The same for a `Milky-Way-like' sample with $M_r = -21 \pm 1$, compared with the relative luminosity function for the Milky Way, with size and magnitude cuts (solid blue line), a magnitude cut only (dashed red line), or no cuts on size or magnitude (dotted magenta line).}
\label{fig08}
\end{figure*}

\subsection{Optimal Weighting}

Splitting the satellite sample into many $\Delta m_r$ magnitude bins greatly reduces the signal-to-noise, particularly for smaller subsamples of the primaries. Due to the drastic reduction in number counts per bin, we run into the problem of extremely noisy background estimates. To overcome this we use a mean background model to determine the relative excess in counts. 

First, we construct an apparent ($r$-band) luminosity function for all the SDSS galaxies in our satellite sample. For a given primary of magnitude $m_{r,{\rm main}}$, we then calculate the fraction of the sample which would lie within a relative offset $\Delta m_r$ of $m_{r,{\rm main}}$.
Finally we multiply this fraction by the {\it total} background counts around the primary. This gives us a smoother estimate of the average background expected in a single $\Delta m_r$ bin.

The S/N in a given bin can still vary strongly from one primary to the next, particularly since our primary galaxies are at different distances, and thus the magnitude limit of the survey will exclude different $\Delta m_r$ bins for different primaries. To produce an optimal estimate of the mean luminosity function, we calculate the average satellite count in each bin using inverse-variance weighting for the contributions from individual primaries. Thus the average count in bin $i$, $\overline{n}_i$ is calculated using:
\begin{equation}
\overline{n}_i = \frac{\sum_{j=1}^n \left(n_{i,j}/\sigma_{i,j}^2\right)}{\sum_{j=1}^n \left(1/\sigma_{i,j}^2\right)}
\end{equation}
where $n_{i,j}$ is the unweighted count for primary $j$ in bin $i$, and $\sigma_{i,j}^2$ is the uncertainty in the counts for that primary and bin. 

\subsection{Comparison to Previous Work}

Figure \ref{fig07} shows the cumulative relative luminosity function in $r$-band magnitude, $N(\Delta m_r)$, for our bright ($M_K \le -23.5$) and faint ($M_K > -23.5$) subsamples (left-hand panel blue and red points respectively). Only satellites between projected radii of 50 and 400\,kpc are included. The errors bars are the uncertainty in the inverse-variance weighted average. As discussed previously, the Milky Way may be intermediate between these two samples; the right-hand panel shows the average luminosity function for the Milky-Way-like sample with $M_r = -21 \pm 1$ (points with error bars). The solid blue line shows the relative luminosity function for known Milky Way satellites, with cuts in size, magnitude, colour and projected radius applied as in our earlier `Local Group comparison' sample (see section 2.2). The dashed red and dotted magenta lines show the relative luminosity function for Milky Way satellites with only a magnitude cut, or with no cuts at all, respectively.

The relative luminosity function has been measured previously around more distant samples of primaries, down to $\Delta m_r \sim 8$.  \cite{Nierenberg2011} measured down to $\Delta m_r = 6 $ around GOODS ellipticals at redshifts 0.1--0.8, for instance; \cite{Lares2011} measured to $\Delta m_r = 7$--8 in SDSS for galaxies at $z = 0.03$--0.1 (i.e. distances of 120--400\,Mpc); \cite{Guo2011} made similar measurements over a wider range of primary luminosity and out to z = 0.5;  \cite{Strigari2012} measured down to $\Delta m_r = 8$, but also obtained upper limits on the satellite abundance down to $\Delta m = 10$. These results are summarized in \cite{Nierenberg2012}, who also show measurements down to $\Delta m_r = 8$ from COSMOS HST images \citep[see also work since by][]{Wang12,Jiang}. A number of studies have also considered separately the abundance of brighter, LMC-like satellites with $\Delta m = $3--5 \cite[e.g.][]{Liu2011,Prescott11,Sales2012,Guo2012}. Below $\Delta m = 8$, the only discoveries outside the Local Group have been serendipitous \citep[e.g.][who identified four individual satellites below $\Delta m = 8$ using narrow-band $H\alpha$ imaging, but were only complete to $\Delta m \sim 5$]{JamesIvory11}. Thus, our result extends most previous measurements of satellite abundance by 4--5 magnitudes further down the luminosity function. 

Previous work reached several general conclusions on the relative luminosity function, as reviewed in \cite{Nierenberg2012}. Satellite counts are reasonably well fit by a Schechter function, with a shallow power-law slope at intermediate magnitudes (e.g. Lares et al. 2011 measure $\alpha = 1.3 \pm 0.2$), which may steepen at the faint end, below $\Delta m$ = 5--6. On average the slope is steeper than those measured for satellites of the Milky Way or M31, with slightly fewer bright satellites above $\Delta m = 6$, and more satellites below this. The normalization depends strongly on primary luminosity, particularly at the bright end where the number of satellites with $\Delta m < 5$ varies by factors of several, e.g. between a sample with $\log M_{*,{\rm main}} = 10.5-11$  and a sample with $\log M_{*,{\rm main}} = 11-11.5$ \citep[][figure 7]{Nierenberg2012}. The variations in slope and normalization at the faint end are less clear. There is a weaker dependence on primary colour or morphology, with red primaries having more satellites at a given $\Delta m$. There appears to be little or no dependence on the redshift of the sample, at least out to $z \lesssim 1$. 

Our results are roughly consistent with these trends, though with some interesting differences. We find less variation in normalization between our bright and faint samples, possibly because they are both at the faint end of the range studied in previous work. Our isolation cuts may also bias our bright sample to lower mean halo masses than purely magnitude-selected samples. Our Milky-Way-like sample is similar to the lowest luminosity bin of \cite{Guo2011}, however, and for it we find results consistent with theirs over the range $\Delta m = 1$--6 (see their figure 7). For our bright sample, the cumulative counts rise as a power law below $\Delta m = 6$, as seen in \cite{Nierenberg2012}. The faint end rise is steeper than seen for the satellites of the Milky Way, once our cuts in size, magnitude and colour are taken into account (solid blue curve in the right-hand panel). The Milky Way also has an excess of bright, Magellanic-Cloud-like systems with $\Delta m < 6$, although this is only significant for the LMC itself after our cuts are taken into account. Intriguingly, however, we find that fainter primaries have a flatter relative luminosity function, and more satellites with $\Delta m \le 4$. This suggests that satellites like the Magellanic Clouds might be somewhat less exceptional if the luminosity and/or mass of the Milky Way had been {\it over}-estimated. We note recent work by \cite{Gonzalez} shows that less massive halos are also more likely to host a close pair of subhalos with separations and velocity differences as small as those of the Magellanic Clouds.

We note however several caveats when comparing our measured luminosity function to that of the Milky Way or M31. First, our cuts in size and magnitude, required to eliminate the large number of background galaxies, will also exclude the most compact dwarfs. We have also cut out the innermost annulus ($R_p = 0$--50\,kpc) to remove contamination from bright globular clusters and HII regions, as distinguishing these from satellites in a systematic way would require detailed photometric and/or morphological analysis. In the case of the Milky Way, this could remove several nearby satellites from our sample, depending on the projection considered. Comparing the three curves in the right-hand panel of Figure \ref{fig07}, we conclude that our final counts may be incomplete by a factor of 2--3 due to the size, magnitude, colour and projected separation cuts. On the other hand, the relative luminosity functions for the Local Group primaries may be slightly incomplete as well, particularly below $\Delta m \sim 10$. Overall, our results provide a measurement of the abundance of extended satellites around galaxies like the Milky Way down to $\Delta m \sim $12--13, that is $M_r = -9$ to $-8$, as well as a lower bound on the total abundance of all satellites. 

\subsection{Spectroscopic Confirmation}

\begin{figure*}
\plottwo{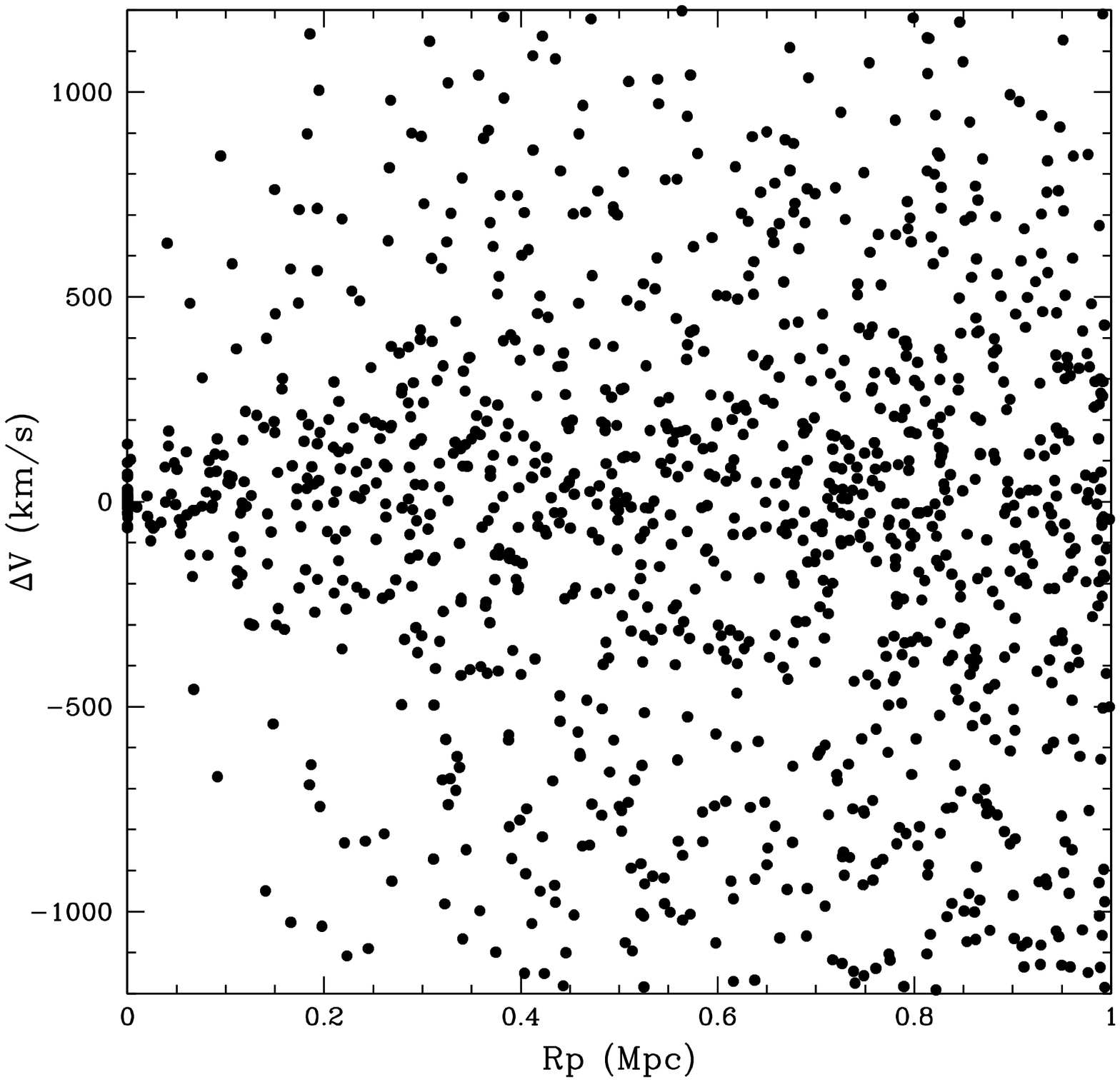}{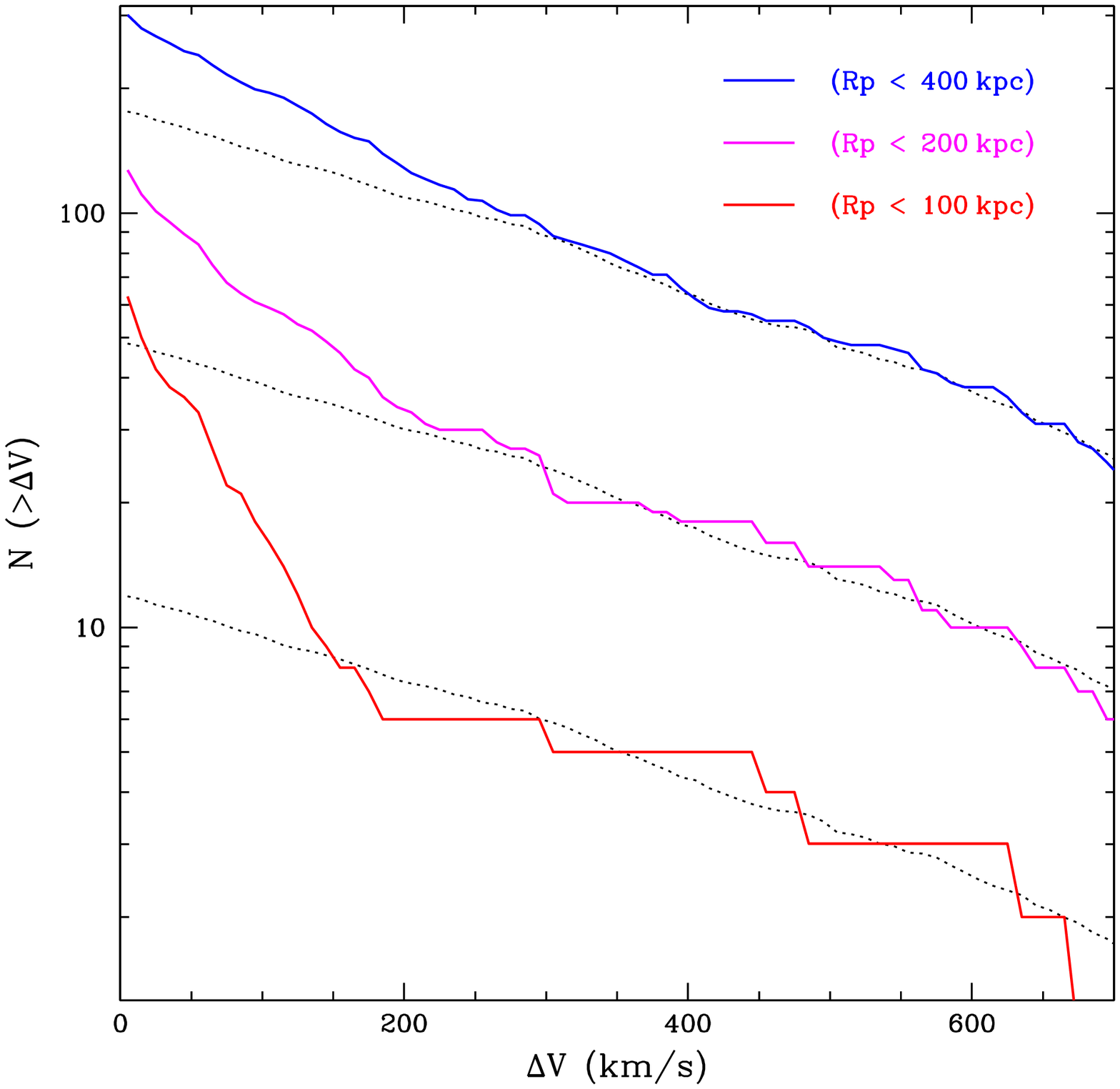}
\caption{{\it Left:} Velocity offset from the primary $\Delta V$ versus projected separation ${\rm R_p}$, for the (brightest) satellites with spectroscopic redshifts. {\it Right:} Number of objects with velocity offsets less than $\Delta V$, in radial ranges ${\rm R_p} = $ 0--100\,kpc (lower red curve), 0--200\,kpc (middle magenta curve), or 0--400\,kpc (upper blue curve). The dotted line shows the distribution of $\Delta V$ at large radii (500\,kpc--1\,Mpc), normalized to match each distribution at large values of $\Delta V$. \label{fig07}}
\end{figure*}

Clustering on the sky gives only statistical evidence for association between faint and bright galaxies. It is natural to want to confirm this for individual objects, using redshifts. Given the magnitude range and spatial distribution of our satellites, this will be logistically challenging. We can obtain redshifts for a small number of the brightest satellites from SDSS itself, however. The limiting  $r$-band magnitude of the SDSS spectroscopic catalogue is 4.5 magnitudes brighter than that of the photometric catalogue, so we will only be able to detect the satellite population down to $\Delta m \sim 7$--8, but we can nonetheless use these bright objects to check for consistency with our clustering results.

Selecting all galaxies with spectra within ${\rm R_p} = 1$\,Mpc of our primaries, and applying the same cuts in size, magnitude and colour to the spectroscopic catalogue as we have the photometric catalogue, we find a total of 7732 objects at all redshifts around 244 of our primaries.
Using primary redshifts from SDSS or the NASA/IPAC Extragalactic Database\footnote{NED -- \url{http://ned.ipac.caltech.edu}}, we calculate the velocity offset $\Delta V = V_{\rm sat} - V_{\rm main}$. The left panel of figure \ref{fig07} shows the distribution of $\Delta V$ versus $R_p$, for the 1159 objects with $|\Delta V| < 1200$ km/s. An over-density of points is clearly visible at  $|\Delta V| < 200$ km/s, corresponding roughly to the velocity dispersion of our primary halos. (We note that the area sampled increases linearly with $R_p$, explaining why there is also a horizontal gradient in density across the plot.) Not all the objects with small $\Delta V$ are satellites, however; this can be inferred from the fact that the overdensity extends all the way out to 1\,Mpc projected separation. In fact, this overdensity at large radii and small $\Delta V$ corresponds to the `two-halo term' in galaxy clustering. The characteristic spatial scale of the two-halo term for our primaries is roughly 4\,Mpc, which translates into a redshift range of $\pm 300$\,km/s. 

Some fraction of the objects with small velocity offsets are genuine satellites. We can see this by combining clustering measurements in both position and velocity space. The right-hand panel of figure \ref{fig07} shows the distribution of velocity offset for objects within three different radial ranges, 0--400, 0--200, and 0--100\,kpc respectively from top to bottom (solid lines). The dotted line shows the distribution of velocity offset for $R_p = $500\,kpc--1\,Mpc, rescaled to the three other distributions so that they agree at large $\Delta V$. The inner radial bins show a clear excess of objects at $\Delta V \lesssim 200$\,km/s relative to the distribution at large projected radii. The excess corresponds to $\sim 100$ satellites within 100 or 200\,kpc projected, or 0.4 satellites per primary galaxy. Our photometric results indicate we should have 1--2 satellites within these radii, down to $\Delta m \sim 12$; given the shape of the relative luminosity function we expect the numbers at $\Delta m = $7--8 to be down by a factor of 3--4 relative to those at $\Delta m = 12$, so the abundance of spectroscopic satellites seems consistent with the clustering measured from the photometric catalogue. This provides an additional consistency check of our previous results, down to $\Delta m_r \sim 8$.

Overall, the distribution of velocity offsets shown in the right-hand panel of figure \ref{fig07} can be understood as the sum of a two-halo term which follows the dotted line, and a one-halo term responsible for the excess at small radii. This illustrates a practical complication in confirming the association of satellites with primaries. Since the spatial extent of the two-halo term along the line of sight translates into a velocity offset comparable to the expected velocity dispersions of our primary halos, separating the one- and two-halo components requires some caution. Even at relatively small projected separation ($R_p < 100$\,kpc), the two-halo term still contributes 15-25\%\ of the excess counts at $\Delta V \le 200$\,km/s.

\section{Summary and Future Prospects}

The faint satellites of bright galaxies such as the Milky Way and M31 provide an important test of galaxy formation models within the CDM framework. Unfortunately, few other samples of faint dwarfs exist comparable to those in the Local Group. Several groups have made progress on this problem recently, identifying the brighter satellites of large numbers of isolated luminous galaxies. Using a very local sample of bright galaxies, we have extended this work 4--5 magnitudes down the luminosity function, to satellites 12--13 magnitudes fainter than their primaries. Although the sample is incomplete, it gives us a partial estimate of the satellite luminosity function down to the bottom of the `classical' dwarf range, $M_r \sim -8$, and rules out a population of faint, moderately extended dwarfs more than 2--3 times larger on average than that currently known in the Milky Way system.

We have shown in particular that it is possible to detect very nearby satellites using relatively shallow photometry. The key to reducing background contamination is a magnitude-dependent size cut, which for very nearby systems selects out relatively low-surface-brightness dwarfs. For nearby, faint galaxies, this is more effective than selection using SDSS photometric redshifts, which can have errors of $\pm0.05$ ($\pm200$\,Mpc) or more. The result is a strong clustering signal, corresponding to a large population of satellites per primary. We note however that the satellite sample is incomplete by construction, excluding the most and least compact objects. Given our cuts, we find $4.65 \pm 0.53$ satellites per primary in our full sample. Although we lack the signal-to-noise to measure an accurate radial profile, the projected distribution is consistent with an isothermal or NFW profile, extending out to $\sim$400\,kpc.

Splitting our primary sample by luminosity, the 25\%\ brightest primaries ($M_{K_s} \le -23.5$) have $\sim$3 times more satellites than the fainter primaries, in a distribution extending out to 500--600\,kpc. Abundance matching and weak gravitational lensing suggest the mean halo mass of our whole primary sample is $\sim 1\times 10^{12}M_\odot$, and the  mean virial radius should be 250\,kpc (though some dynamical studies of satellites give a slightly higher mass estimate -- cf.~\citealt{More}, as discussed in \citealt{Leauthaud}). The bright subsample should occupy halos 3--4 times more massive than those of the fainter primaries, with a virial radius of $\sim$400\,kpc, although our isolation criteria may bias the mean mass of the sample slightly. Given these mass estimates, our satellite detections extend out to 1--1.5 virial radii in projection, and satellite abundance varies approximately linearly with halo mass. The former result is consistent with expectations from simulations; the linear dependence of abundance on halo mass is probably an accidental consequence of the halo mass range sampled. Over a larger range the relationship between stellar and halo mass ratio changes slope, affecting satellite abundance \citep[e.g.][]{Sales11}. 

Splitting the sample by morphology, the ellipticals appear to have $85\%\ \pm 35\%$ more satellites than spirals, clustered on a larger spatial scale of 500--600\,kpc, even though the average primary luminosities of the two subsamples are equal. Thus, we find evidence for a morphological dependence of the stellar luminosity to halo mass ratio, albeit at marginal statistical significance. This could reflect differences in  the mass-to-light ratio of the stellar populations, or possibly a more fundamental difference in the stellar-to-halo mass ratio. 

One slight puzzle is the colour range of our satellite population; the majority of the satellites appear to be blue $(g-r < 0.6)$ as opposed to red $(g -r \ge 0.6)$. \cite{Guo2011} found similar results, but only for brighter satellites ($\Delta m \le 6$). In subsequent work \citep{Guo2012} they also showed that red satellites cluster more strongly than blue satellites, so it may be that a larger fraction of red satellites are excluded by our inner radial cut at 50\,kpc. Our size cut probably has the largest effect on the colour distribution, however; we expect that a looser cut on size would reveal more of the faint red dwarf spheroidals common in the Local Group. Nonetheless, understanding the colour dependence of the abundance and spatial distribution of satellites remains problematic, as discussed in the most recent work by \cite{Guo2013}. 

Beyond this statistical detection of the satellite population, confirming the identity of individual satellites will be challenging. While satellites could in principle be identified spectroscopically, the field sizes and magnitude limits make this prohibitively expensive for a large sample of primaries.  Furthermore, because the two-halo clustering scale maps onto a line-of-sight velocity range comparable to the internal velocity dispersion of a bright galaxy halo, separating satellites from interlopers is not completely trivial even with full velocity information, as discussed in the preceding section. 

It would be easy, however, to obtain deeper photometry on many or all of our target fields. These extend our 1--2 degrees around each primary, and are well-suited to large-area cameras such as MegaCam \citep{Megacam} or Hyper Suprime-Cam \citep{Hypersuprimecam}. Surveys with these instruments could extend the clustering measurement around a nearby sample like the Atlas3D sample down to larger $\Delta m$, or they could compile better statistics on a larger sample of primaries 2--3 times further away. The only uncertainty in making projections for such follow-up work is to determine the efficiency of the size cut for different distance ranges and magnitude limits. Larger samples would be invaluable, however, to clarify the detailed radial distribution, anisotropy, and colour distribution of faint satellites, halo-to-halo variations, and the link between faint satellites and primary properties. One intriguing idea is to use faint satellite counts as a proxy for halo mass, since simulations suggest the two should correlate reasonably well over certain ranges of primary and satellite luminosity
\citep[e.g.][]{Nickerson}. While the potential accuracy of this method is unclear, it might just be possible to test it directly at slightly larger distances than those considered here by combining clustering measurements with optimally weighted galaxy-galaxy lensing.

Larger samples of faint satellites would provide useful input to models of galaxy formation as well. A particularly important goal, as discussed in the introduction, is to understand what produces the cutoff in the efficiency of galaxy formation. If galaxy formation is stochastic on small scales, for instance, as suggested both by spatial clustering \citep{Taylor04,Kravtsov04b}, and by rotation curves or velocity dispersion profiles \citep{BK11}, what is the hidden variable, if any, that determined which subhalos form stars? The Local Group doesn't have enough faint objects to distinguish easily between alternate models, and may not be representative in any case, so finding well-sampled Local Group analogues is essential in the longer term.

\acknowledgments

We thank Jonathan Grossauer, Gretchen Harris, Michael Balogh, Anna Nierenberg, Michael Drinkwater, Anne-Marie Weijmans, and the participants of the KITP Workshop ``First Galaxies and Faint Dwarfs", for helpful discussions. We also wish to thank the Atlas3D and SDSS teams for making their catalogues available to the community. This work was supported by a Discovery Grant to JET from the Natural Sciences and Engineering Research Council (NSERC) of Canada. 

This project has made use of data from the Sloan Digital Sky Survey III (SDSS-III). Funding for SDSS-III has been provided by the Alfred P. Sloan Foundation, the Participating Institutions, the National Science Foundation, and the U.S. Department of Energy Office of Science. SDSS-III is managed by the Astrophysical Research Consortium for the Participating Institutions of the SDSS-III Collaboration -- see the SDSS-III web site http://www.sdss3.org for a full list of institutions. 


\bibliographystyle{apj}
\bibliography{PaperI.v1}

\begin{thebibliography}{76}
\expandafter\ifx\csname natexlab\endcsname\relax\def\natexlab#1{#1}\fi

\bibitem[{{Barkana} \& {Loeb}(1999)}]{BarkanaLoeb}
{Barkana}, R., \& {Loeb}, A. 1999, \apj, 523, 54

\bibitem[{{Behroozi} {et~al.}(2012){Behroozi}, {Wechsler}, \&
  {Conroy}}]{Behroozi12}
{Behroozi}, P.~S., {Wechsler}, R.~H., \& {Conroy}, C. 2012, ArXiv e-prints

\bibitem[{{Belokurov} {et~al.}(2006){Belokurov}, {Zucker}, {Evans},
  {Wilkinson}, {Irwin}, {Hodgkin}, {Bramich}, {Irwin}, {Gilmore}, {Willman},
  {Vidrih}, {Newberg}, {Wyse}, {Fellhauer}, {Hewett}, {Cole}, {Bell}, {Beers},
  {Rockosi}, {Yanny}, {Grebel}, {Schneider}, {Lupton}, {Barentine},
  {Brewington}, {Brinkmann}, {Harvanek}, {Kleinman}, {Krzesinski}, {Long},
  {Nitta}, {Smith}, \& {Snedden}}]{Belokurov06}
{Belokurov}, V., {et~al.} 2006, \apjl, 647, L111

\bibitem[{{Benson} {et~al.}(2003){Benson}, {Bower}, {Frenk}, {Lacey}, {Baugh},
  \& {Cole}}]{Benson03}
{Benson}, A.~J., {Bower}, R.~G., {Frenk}, C.~S., {Lacey}, C.~G., {Baugh},
  C.~M., \& {Cole}, S. 2003, \apj, 599, 38

\bibitem[{{Benson} {et~al.}(2002){Benson}, {Frenk}, {Lacey}, {Baugh}, \&
  {Cole}}]{Benson02}
{Benson}, A.~J., {Frenk}, C.~S., {Lacey}, C.~G., {Baugh}, C.~M., \& {Cole}, S.
  2002, \mnras, 333, 177

\bibitem[{{Blanton} \& {Roweis}(2007)}]{Blanton}
{Blanton}, M.~R., \& {Roweis}, S. 2007, \aj, 133, 734

\bibitem[{{Boulade} {et~al.}(2003){Boulade}, {Charlot}, {Abbon}, {Aune},
  {Borgeaud}, {Carton}, {Carty}, {Da Costa}, {Deschamps}, {Desforge},
  {Eppell{\'e}}, {Gallais}, {Gosset}, {Granelli}, {Gros}, {de Kat}, {Loiseau},
  {Ritou}, {Rouss{\'e}}, {Starzynski}, {Vignal}, \& {Vigroux}}]{Megacam}
{Boulade}, O., {et~al.} 2003, in Society of Photo-Optical Instrumentation
  Engineers (SPIE) Conference Series, Vol. 4841, Society of Photo-Optical
  Instrumentation Engineers (SPIE) Conference Series, ed. M.~{Iye} \& A.~F.~M.
  {Moorwood}, 72--81

\bibitem[{{Boylan-Kolchin} {et~al.}(2011){Boylan-Kolchin}, {Bullock}, \&
  {Kaplinghat}}]{BK11}
{Boylan-Kolchin}, M., {Bullock}, J.~S., \& {Kaplinghat}, M. 2011, \mnras, 415,
  L40

\bibitem[{{Boylan-Kolchin} {et~al.}(2010){Boylan-Kolchin}, {Springel}, {White},
  \& {Jenkins}}]{BK10}
{Boylan-Kolchin}, M., {Springel}, V., {White}, S.~D.~M., \& {Jenkins}, A. 2010,
  \mnras, 406, 896

\bibitem[{{Bullock} {et~al.}(2000){Bullock}, {Kravtsov}, \&
  {Weinberg}}]{Bullock00}
{Bullock}, J.~S., {Kravtsov}, A.~V., \& {Weinberg}, D.~H. 2000, \apj, 539, 517

\bibitem[{{Busha} {et~al.}(2011){Busha}, {Wechsler}, {Behroozi}, {Gerke},
  {Klypin}, \& {Primack}}]{Busha11}
{Busha}, M.~T., {Wechsler}, R.~H., {Behroozi}, P.~S., {Gerke}, B.~F., {Klypin},
  A.~A., \& {Primack}, J.~R. 2011, \apj, 743, 117

\bibitem[{{Cappellari} {et~al.}(2011){Cappellari}, {Emsellem}, {Krajnovi{\'c}},
  {McDermid}, {Scott}, {Verdoes Kleijn}, {Young}, {Alatalo}, {Bacon}, {Blitz},
  {Bois}, {Bournaud}, {Bureau}, {Davies}, {Davis}, {de Zeeuw}, {Duc},
  {Khochfar}, {Kuntschner}, {Lablanche}, {Morganti}, {Naab}, {Oosterloo},
  {Sarzi}, {Serra}, \& {Weijmans}}]{Atlas3D}
{Cappellari}, M., {et~al.} 2011, \mnras, 413, 813

\bibitem[{{Carlberg} {et~al.}(2009){Carlberg}, {Sullivan}, \& {Le
  Borgne}}]{Carlberg}
{Carlberg}, R.~G., {Sullivan}, M., \& {Le Borgne}, D. 2009, \apj, 694, 1131

\bibitem[{{Chen}(2008)}]{Chen08}
{Chen}, J. 2008, \aap, 484, 347

\bibitem[{{Chen} {et~al.}(2006){Chen}, {Kravtsov}, {Prada}, {Sheldon},
  {Klypin}, {Blanton}, {Brinkmann}, \& {Thakar}}]{Chen2006}
{Chen}, J., {Kravtsov}, A.~V., {Prada}, F., {Sheldon}, E.~S., {Klypin}, A.~A.,
  {Blanton}, M.~R., {Brinkmann}, J., \& {Thakar}, A.~R. 2006, \apj, 647, 86

\bibitem[{{de Vaucouleurs} {et~al.}(1991){de Vaucouleurs}, {de Vaucouleurs},
  {Corwin}, {Buta}, {Paturel}, \& {Fouqu{\'e}}}]{RC3}
{de Vaucouleurs}, G., {de Vaucouleurs}, A., {Corwin}, Jr., H.~G., {Buta},
  R.~J., {Paturel}, G., \& {Fouqu{\'e}}, P. 1991, {Third Reference Catalogue of
  Bright Galaxies.} (Springer)

\bibitem[{{Dekel} \& {Silk}(1986)}]{DekelSilk}
{Dekel}, A., \& {Silk}, J. 1986, \apj, 303, 39

\bibitem[{{Efstathiou}(1992)}]{Efstathiou92}
{Efstathiou}, G. 1992, \mnras, 256, 43P

\bibitem[{{Fukugita} {et~al.}(1995){Fukugita}, {Shimasaku}, \&
  {Ichikawa}}]{Fukugita}
{Fukugita}, M., {Shimasaku}, K., \& {Ichikawa}, T. 1995, \pasp, 107, 945

\bibitem[{{Gnedin} \& {Kravtsov}(2006)}]{Gnedin06}
{Gnedin}, N.~Y., \& {Kravtsov}, A.~V. 2006, \apj, 645, 1054

\bibitem[{{Gonzalez} {et~al.}(2013){Gonzalez}, {Kravtsov}, \&
  {Gnedin}}]{Gonzalez}
{Gonzalez}, R.~E., {Kravtsov}, A.~V., \& {Gnedin}, N.~Y. 2013, ArXiv e-prints

\bibitem[{{Governato} {et~al.}(2010){Governato}, {Brook}, {Mayer}, {Brooks},
  {Rhee}, {Wadsley}, {Jonsson}, {Willman}, {Stinson}, {Quinn}, \&
  {Madau}}]{Governato10}
{Governato}, F., {et~al.} 2010, \nat, 463, 203

\bibitem[{{Gunn} \& {Gott}(1972)}]{GunnGott}
{Gunn}, J.~E., \& {Gott}, III, J.~R. 1972, \apj, 176, 1

\bibitem[{{Guo} {et~al.}(2011){Guo}, {Cole}, {Eke}, \& {Frenk}}]{Guo2011}
{Guo}, Q., {Cole}, S., {Eke}, V., \& {Frenk}, C. 2011, \mnras, 417, 370

\bibitem[{{Guo} {et~al.}(2012){Guo}, {Cole}, {Eke}, \& {Frenk}}]{Guo2012}
---. 2012, \mnras, 427, 428

\bibitem[{{Guo} {et~al.}(2013){Guo}, {Cole}, {Eke}, {Frenk}, \&
  {Helly}}]{Guo2013}
{Guo}, Q., {Cole}, S., {Eke}, V., {Frenk}, C., \& {Helly}, J. 2013, ArXiv
  e-prints

\bibitem[{{Guo} {et~al.}(2010){Guo}, {White}, {Li}, \&
  {Boylan-Kolchin}}]{Guo_abundance}
{Guo}, Q., {White}, S., {Li}, C., \& {Boylan-Kolchin}, M. 2010, \mnras, 404,
  1111

\bibitem[{{Ibata} {et~al.}(2013){Ibata}, {Lewis}, {Conn}, {Irwin},
  {McConnachie}, {Chapman}, {Collins}, {Fardal}, {Ferguson}, {Ibata}, {Mackey},
  {Martin}, {Navarro}, {Rich}, {Valls-Gabaud}, \& {Widrow}}]{Ibata2013}
{Ibata}, R.~A., {et~al.} 2013, \nat, 493, 62

\bibitem[{{Jackson} {et~al.}(2010){Jackson}, {Bryan}, {Mao}, \& {Li}}]{Jackson}
{Jackson}, N., {Bryan}, S.~E., {Mao}, S., \& {Li}, C. 2010, \mnras, 403, 826

\bibitem[{{James} \& {Ivory}(2011)}]{JamesIvory11}
{James}, P.~A., \& {Ivory}, C.~F. 2011, \mnras, 411, 495

\bibitem[{{Jiang} {et~al.}(2012){Jiang}, {Jing}, \& {Li}}]{Jiang}
{Jiang}, C.~Y., {Jing}, Y.~P., \& {Li}, C. 2012, \apj, 760, 16

\bibitem[{{Karachentsev}(2005)}]{Karachentsev05}
{Karachentsev}, I.~D. 2005, \aj, 129, 178

\bibitem[{{Kauffmann} {et~al.}(1993){Kauffmann}, {White}, \&
  {Guiderdoni}}]{KWG}
{Kauffmann}, G., {White}, S.~D.~M., \& {Guiderdoni}, B. 1993, \mnras, 264, 201

\bibitem[{{Klypin} {et~al.}(1999){Klypin}, {Kravtsov}, {Valenzuela}, \&
  {Prada}}]{Klypin99}
{Klypin}, A., {Kravtsov}, A.~V., {Valenzuela}, O., \& {Prada}, F. 1999, \apj,
  522, 82

\bibitem[{{Koposov} {et~al.}(2008){Koposov}, {Belokurov}, {Evans}, {Hewett},
  {Irwin}, {Gilmore}, {Zucker}, {Rix}, {Fellhauer}, {Bell}, \&
  {Glushkova}}]{Koposov08}
{Koposov}, S., {et~al.} 2008, \apj, 686, 279

\bibitem[{{Kravtsov} {et~al.}(2004){Kravtsov}, {Gnedin}, \&
  {Klypin}}]{Kravtsov04b}
{Kravtsov}, A.~V., {Gnedin}, O.~Y., \& {Klypin}, A.~A. 2004, \apj, 609, 482

\bibitem[{{Lares} {et~al.}(2011){Lares}, {Lambas}, \&
  {Dom{\'{\i}}nguez}}]{Lares2011}
{Lares}, M., {Lambas}, D.~G., \& {Dom{\'{\i}}nguez}, M.~J. 2011, \aj, 142, 13

\bibitem[{{Leauthaud} {et~al.}(2012){Leauthaud}, {Tinker}, {Bundy}, {Behroozi},
  {Massey}, {Rhodes}, {George}, {Kneib}, {Benson}, {Wechsler}, {Busha},
  {Capak}, {Cort{\^e}s}, {Ilbert}, {Koekemoer}, {Le F{\`e}vre}, {Lilly},
  {McCracken}, {Salvato}, {Schrabback}, {Scoville}, {Smith}, \&
  {Taylor}}]{Leauthaud}
{Leauthaud}, A., {et~al.} 2012, \apj, 744, 159

\bibitem[{{Liu} {et~al.}(2011){Liu}, {Gerke}, {Wechsler}, {Behroozi}, \&
  {Busha}}]{Liu2011}
{Liu}, L., {Gerke}, B.~F., {Wechsler}, R.~H., {Behroozi}, P.~S., \& {Busha},
  M.~T. 2011, \apj, 733, 62

\bibitem[{{{\L}okas} {et~al.}(2012){{\L}okas}, {Kazantzidis}, \&
  {Mayer}}]{Lokas}
{{\L}okas}, E.~L., {Kazantzidis}, S., \& {Mayer}, L. 2012, \apjl, 751, L15

\bibitem[{{Lorrimer} {et~al.}(1994){Lorrimer}, {Frenk}, {Smith}, {White}, \&
  {Zaritsky}}]{Lorrimer}
{Lorrimer}, S.~J., {Frenk}, C.~S., {Smith}, R.~M., {White}, S.~D.~M., \&
  {Zaritsky}, D. 1994, \mnras, 269, 696

\bibitem[{{Lynden-Bell} \& {Lynden-Bell}(1995)}]{LyndenBell}
{Lynden-Bell}, D., \& {Lynden-Bell}, R.~M. 1995, \mnras, 275, 429

\bibitem[{{Mashchenko} {et~al.}(2008){Mashchenko}, {Wadsley}, \&
  {Couchman}}]{Maschenko08}
{Mashchenko}, S., {Wadsley}, J., \& {Couchman}, H.~M.~P. 2008, Science, 319,
  174

\bibitem[{{Mayer} {et~al.}(2006){Mayer}, {Mastropietro}, {Wadsley}, {Stadel},
  \& {Moore}}]{Mayer06}
{Mayer}, L., {Mastropietro}, C., {Wadsley}, J., {Stadel}, J., \& {Moore}, B.
  2006, \mnras, 369, 1021

\bibitem[{{McConnachie}(2012)}]{McConnachierev}
{McConnachie}, A.~W. 2012, \aj, 144, 4

\bibitem[{{Moore} {et~al.}(1999){Moore}, {Ghigna}, {Governato}, {Lake},
  {Quinn}, {Stadel}, \& {Tozzi}}]{Moore99}
{Moore}, B., {Ghigna}, S., {Governato}, F., {Lake}, G., {Quinn}, T., {Stadel},
  J., \& {Tozzi}, P. 1999, \apjl, 524, L19

\bibitem[{{Moore} {et~al.}(1996){Moore}, {Katz}, {Lake}, {Dressler}, \&
  {Oemler}}]{Moore96}
{Moore}, B., {Katz}, N., {Lake}, G., {Dressler}, A., \& {Oemler}, A. 1996,
  \nat, 379, 613

\bibitem[{{More} {et~al.}(2011){More}, {van den Bosch}, {Cacciato}, {Skibba},
  {Mo}, \& {Yang}}]{More}
{More}, S., {van den Bosch}, F.~C., {Cacciato}, M., {Skibba}, R., {Mo}, H.~J.,
  \& {Yang}, X. 2011, \mnras, 410, 210

\bibitem[{{Nichols} \& {Bland-Hawthorn}(2011)}]{Nichols11}
{Nichols}, M., \& {Bland-Hawthorn}, J. 2011, \apj, 732, 17

\bibitem[{{Nichols} {et~al.}(2011){Nichols}, {Colless}, {Colless}, \&
  {Bland-Hawthorn}}]{Nichols11b}
{Nichols}, M., {Colless}, J., {Colless}, M., \& {Bland-Hawthorn}, J. 2011,
  \apj, 742, 110

\bibitem[{{Nickerson} {et~al.}(2013){Nickerson}, {Stinson}, {Couchman},
  {Bailin}, \& {Wadsley}}]{Nickerson}
{Nickerson}, S., {Stinson}, G., {Couchman}, H.~M.~P., {Bailin}, J., \&
  {Wadsley}, J. 2013, \mnras, 429, 452

\bibitem[{{Nierenberg} {et~al.}(2011){Nierenberg}, {Auger}, {Treu}, {Marshall},
  \& {Fassnacht}}]{Nierenberg2011}
{Nierenberg}, A.~M., {Auger}, M.~W., {Treu}, T., {Marshall}, P.~J., \&
  {Fassnacht}, C.~D. 2011, \apj, 731, 44

\bibitem[{{Nierenberg} {et~al.}(2012){Nierenberg}, {Auger}, {Treu}, {Marshall},
  {Fassnacht}, \& {Busha}}]{Nierenberg2012}
{Nierenberg}, A.~M., {Auger}, M.~W., {Treu}, T., {Marshall}, P.~J.,
  {Fassnacht}, C.~D., \& {Busha}, M.~T. 2012, \apj, 752, 99

\bibitem[{{Prescott} {et~al.}(2011){Prescott}, {Baldry}, {James}, {Bamford},
  {Bland-Hawthorn}, {Brough}, {Brown}, {Cameron}, {Conselice}, {Croom},
  {Driver}, {Frenk}, {Gunawardhana}, {Hill}, {Hopkins}, {Jones}, {Kelvin},
  {Kuijken}, {Liske}, {Loveday}, {Nichol}, {Norberg}, {Parkinson}, {Peacock},
  {Phillipps}, {Pimbblet}, {Popescu}, {Robotham}, {Sharp}, {Sutherland},
  {Taylor}, {Tuffs}, {van Kampen}, \& {Wijesinghe}}]{Prescott11}
{Prescott}, M., {et~al.} 2011, \mnras, 417, 1374

\bibitem[{{Rees} \& {Ostriker}(1977)}]{ReesOstriker}
{Rees}, M.~J., \& {Ostriker}, J.~P. 1977, \mnras, 179, 541

\bibitem[{{Robotham} {et~al.}(2012){Robotham}, {Baldry}, {Bland-Hawthorn},
  {Driver}, {Loveday}, {Norberg}, {Bauer}, {Bekki}, {Brough}, {Brown},
  {Graham}, {Hopkins}, {Phillipps}, {Power}, {Sansom}, \&
  {Staveley-Smith}}]{Robotham12}
{Robotham}, A.~S.~G., {et~al.} 2012, \mnras, 424, 1448

\bibitem[{{Sales} {et~al.}(2011){Sales}, {Navarro}, {Cooper}, {White}, {Frenk},
  \& {Helmi}}]{Sales11}
{Sales}, L.~V., {Navarro}, J.~F., {Cooper}, A.~P., {White}, S.~D.~M., {Frenk},
  C.~S., \& {Helmi}, A. 2011, \mnras, 418, 648

\bibitem[{{Sales} {et~al.}(2013){Sales}, {Wang}, {White}, \&
  {Navarro}}]{Sales2012}
{Sales}, L.~V., {Wang}, W., {White}, S.~D.~M., \& {Navarro}, J.~F. 2013,
  \mnras, 428, 573

\bibitem[{{Silk}(1977)}]{Silk}
{Silk}, J. 1977, \apj, 211, 638

\bibitem[{{Skrutskie} {et~al.}(2006){Skrutskie}, {Cutri}, {Stiening},
  {Weinberg}, {Schneider}, {Carpenter}, {Beichman}, {Capps}, {Chester},
  {Elias}, {Huchra}, {Liebert}, {Lonsdale}, {Monet}, {Price}, {Seitzer},
  {Jarrett}, {Kirkpatrick}, {Gizis}, {Howard}, {Evans}, {Fowler}, {Fullmer},
  {Hurt}, {Light}, {Kopan}, {Marsh}, {McCallon}, {Tam}, {Van Dyk}, \&
  {Wheelock}}]{2MASS}
{Skrutskie}, M.~F., {et~al.} 2006, \aj, 131, 1163

\bibitem[{{Strauss} {et~al.}(2002){Strauss}, {Weinberg}, {Lupton}, {Narayanan},
  {Annis}, {Bernardi}, {Blanton}, {Burles}, {Connolly}, {Dalcanton}, {Doi},
  {Eisenstein}, {Frieman}, {Fukugita}, {Gunn}, {Ivezi{\'c}}, {Kent}, {Kim},
  {Knapp}, {Kron}, {Munn}, {Newberg}, {Nichol}, {Okamura}, {Quinn}, {Richmond},
  {Schlegel}, {Shimasaku}, {SubbaRao}, {Szalay}, {Vanden Berk}, {Vogeley},
  {Yanny}, {Yasuda}, {York}, \& {Zehavi}}]{Strauss02}
{Strauss}, M.~A., {et~al.} 2002, \aj, 124, 1810

\bibitem[{{Strigari} \& {Wechsler}(2012)}]{Strigari2012}
{Strigari}, L.~E., \& {Wechsler}, R.~H. 2012, \apj, 749, 75

\bibitem[{{Takada}(2010)}]{Hypersuprimecam}
{Takada}, M. 2010, in American Institute of Physics Conference Series, Vol.
  1279, American Institute of Physics Conference Series, ed. N.~{Kawai} \&
  S.~{Nagataki}, 120--127

\bibitem[{{Tal} {et~al.}(2012){Tal}, {Wake}, \& {van Dokkum}}]{Tal}
{Tal}, T., {Wake}, D.~A., \& {van Dokkum}, P.~G. 2012, \apjl, 751, L5

\bibitem[{{Taylor} \& {Babul}(2001)}]{TB01}
{Taylor}, J.~E., \& {Babul}, A. 2001, \apj, 559, 716

\bibitem[{{Taylor} {et~al.}(2004){Taylor}, {Babul}, \& {Silk}}]{Taylor04}
{Taylor}, J.~E., {Babul}, A., \& {Silk}, J. 2004, in Astronomical Society of
  the Pacific Conference Series, Vol. 327, Satellites and Tidal Streams, ed.
  F.~{Prada}, D.~{Martinez Delgado}, \& T.~J. {Mahoney}, 205

\bibitem[{{Tollerud} {et~al.}(2011){Tollerud}, {Boylan-Kolchin}, {Barton},
  {Bullock}, \& {Trinh}}]{Tollerud11}
{Tollerud}, E.~J., {Boylan-Kolchin}, M., {Barton}, E.~J., {Bullock}, J.~S., \&
  {Trinh}, C.~Q. 2011, \apj, 738, 102

\bibitem[{{Tollerud} {et~al.}(2008){Tollerud}, {Bullock}, {Strigari}, \&
  {Willman}}]{Tollerud08}
{Tollerud}, E.~J., {Bullock}, J.~S., {Strigari}, L.~E., \& {Willman}, B. 2008,
  \apj, 688, 277

\bibitem[{{van den Bergh}(2000)}]{vandenBergh}
{van den Bergh}, S. 2000, {The Galaxies of the Local Group} (Cambridge)

\bibitem[{{Wang} \& {White}(2012)}]{Wang12}
{Wang}, W., \& {White}, S.~D.~M. 2012, \mnras, 424, 2574

\bibitem[{{White} \& {Frenk}(1991)}]{WhiteFrenk}
{White}, S.~D.~M., \& {Frenk}, C.~S. 1991, \apj, 379, 52

\bibitem[{{White} \& {Rees}(1978)}]{WhiteRees}
{White}, S.~D.~M., \& {Rees}, M.~J. 1978, \mnras, 183, 341

\bibitem[{{Williams} {et~al.}(2009){Williams}, {Bureau}, \&
  {Cappellari}}]{WIlliams}
{Williams}, M.~J., {Bureau}, M., \& {Cappellari}, M. 2009, \mnras, 400, 1665

\bibitem[{{Willman} {et~al.}(2005){Willman}, {Dalcanton}, {Martinez-Delgado},
  {West}, {Blanton}, {Hogg}, {Barentine}, {Brewington}, {Harvanek}, {Kleinman},
  {Krzesinski}, {Long}, {Neilsen}, {Nitta}, \& {Snedden}}]{Willman05}
{Willman}, B., {et~al.} 2005, \apjl, 626, L85

\bibitem[{{York} {et~al.}(2000){York}, {Adelman}, {Anderson}, {Anderson},
  {Annis}, {Bahcall}, {Bakken}, {Barkhouser}, {Bastian}, {Berman}, {Boroski},
  {Bracker}, {Briegel}, {Briggs}, {Brinkmann}, {Brunner}, {Burles}, {Carey},
  {Carr}, {Castander}, {Chen}, {Colestock}, {Connolly}, {Crocker}, {Csabai},
  {Czarapata}, {Davis}, {Doi}, {Dombeck}, {Eisenstein}, {Ellman}, {Elms},
  {Evans}, {Fan}, {Federwitz}, {Fiscelli}, {Friedman}, {Frieman}, {Fukugita},
  {Gillespie}, {Gunn}, {Gurbani}, {de Haas}, {Haldeman}, {Harris}, {Hayes},
  {Heckman}, {Hennessy}, {Hindsley}, {Holm}, {Holmgren}, {Huang}, {Hull},
  {Husby}, {Ichikawa}, {Ichikawa}, {Ivezi{\'c}}, {Kent}, {Kim}, {Kinney},
  {Klaene}, {Kleinman}, {Kleinman}, {Knapp}, {Korienek}, {Kron}, {Kunszt},
  {Lamb}, {Lee}, {Leger}, {Limmongkol}, {Lindenmeyer}, {Long}, {Loomis},
  {Loveday}, {Lucinio}, {Lupton}, {MacKinnon}, {Mannery}, {Mantsch}, {Margon},
  {McGehee}, {McKay}, {Meiksin}, {Merelli}, {Monet}, {Munn}, {Narayanan},
  {Nash}, {Neilsen}, {Neswold}, {Newberg}, {Nichol}, {Nicinski}, {Nonino},
  {Okada}, {Okamura}, {Ostriker}, {Owen}, {Pauls}, {Peoples}, {Peterson},
  {Petravick}, {Pier}, {Pope}, {Pordes}, {Prosapio}, {Rechenmacher}, {Quinn},
  {Richards}, {Richmond}, {Rivetta}, {Rockosi}, {Ruthmansdorfer}, {Sandford},
  {Schlegel}, {Schneider}, {Sekiguchi}, {Sergey}, {Shimasaku}, {Siegmund},
  {Smee}, {Smith}, {Snedden}, {Stone}, {Stoughton}, {Strauss}, {Stubbs},
  {SubbaRao}, {Szalay}, {Szapudi}, {Szokoly}, {Thakar}, {Tremonti}, {Tucker},
  {Uomoto}, {Vanden Berk}, {Vogeley}, {Waddell}, {Wang}, {Watanabe},
  {Weinberg}, {Yanny}, {Yasuda}, \& {SDSS Collaboration}}]{SDSS}
{York}, D.~G., {et~al.} 2000, \aj, 120, 1579

\bibitem[{{Zucker} {et~al.}(2006){Zucker}, {Belokurov}, {Evans}, {Kleyna},
  {Irwin}, {Wilkinson}, {Fellhauer}, {Bramich}, {Gilmore}, {Newberg}, {Yanny},
  {Smith}, {Hewett}, {Bell}, {Rix}, {Gnedin}, {Vidrih}, {Wyse}, {Willman},
  {Grebel}, {Schneider}, {Beers}, {Kniazev}, {Barentine}, {Brewington},
  {Brinkmann}, {Harvanek}, {Kleinman}, {Krzesinski}, {Long}, {Nitta}, \&
  {Snedden}}]{Zucker06}
{Zucker}, D.~B., {et~al.} 2006, \apjl, 650, L41

\end{thebibliography}

\end{document}